 \documentclass[final,5p,times,twocolumn]{elsarticle}

\usepackage{CJKutf8}
\usepackage{listings}
\usepackage{graphicx}
\usepackage{epsfig}
\usepackage{amsmath,amssymb}
\usepackage{bm}
\usepackage{hyperref}
\usepackage{color}
\usepackage{float}
\usepackage{dcolumn}
\usepackage{multirow}
\usepackage{mathrsfs}
\usepackage{physics}
\usepackage[OT1]{fontenc}
\usepackage{outline}
\usepackage{tikz,tkz-euclide}
\usetikzlibrary{arrows,calc,patterns}
\usetikzlibrary{positioning}

\usepackage{amssymb}

\usepackage[OT1]{fontenc}

\newcommand{\HIT}{\textsc{hit3d}}

\newcounter{bla}

\journal{Computer Physics Communications}

\begin{document}

\begin{frontmatter}



\title{A program for 3D nuclear static and time-dependent density-functional theory 
       with full Skyrme energy density functional: \HIT}


\author[a]{Yue Shi\corref{author}}
\author[b]{Paul D. Stevenson}
\author[c,d,e]{Nobuo Hinohara}

\cortext[author] {Corresponding author.\\\textit{E-mail address:} yueshi@hit.edu.cn}
\address[a]{Department of Physics, Harbin Institute of Technology, Harbin 150001, People's Republic of China}
\address[b]{School of Mathematics and Physics, University of Surrey, Guildford, GU2 7XH, United Kindom}
\address[c]{Center for Computational Sciences, University of Tsukuba, Tsukuba 305-8577, Japan}
\address[d]{Faculty of Pure and Applied Sciences, University of Tsukuba, Tsukuba, 35-8571 Japan}
\address[e]{Facility for Rare Isotope Beams, Michigan State University, East Lansing, Michigan 48824, USA}

\begin{abstract}

	This work presents a computer program that performs 
	symmetry-unrestricted 3D nuclear time-dependent density function theory
	(DFT) calculations. The program features the augmented Lagrangian
	constraint in the static calculation. This allows for the calculation
	of the potential energy surface. In addition, the code includes the
	full energy density functionals derived from the Skyrme interaction,
	meaning that the time-even and time-odd tensor parts are included for
	the time-dependent calculations. The results of the \HIT~code are
	carefully compared with the Sky3D and Ev8 programs. The testing cases
	include unconstrained DFT calculations for doubly magic nuclei, the
	constrained DFT + BCS calculations for medium-heavy nucleus $^{110}$Mo,
	and the dynamic applications for harmonic vibration and nuclear
	reactions.

\end{abstract}


\end{frontmatter}



{\bf PROGRAM SUMMARY}

\begin{small}
\noindent
{\em Program Title: \HIT}                                          \\
{\em CPC Library link to program files:}                           \\
{\em Code Ocean capsule:}                                           \\
{\em Licensing provisions: CC0 1.0}                                 \\
{\em Programming language: C/C++}                                   \\
{\em Supplementary material:}                                 \\
{\em Nature of problem: The program performs symmetry-unrestricted
	time-dependent (TD) density-functional theory (DFT) calculations for
	finite nuclei. For the static calculations, it performs precise
	constraint calculations using the augmented Lagrangian method. For
	vibrational calculations, the program performs Fourier analysis of 
	the multipole moments as functions of time. The energy-weighted sum rule
	obtained in this manner can be compared with those obtained using the
	ground state densities. In addition, the program performs collision
	calculations.}   \\
{\em Solution method: The single-particle wave functions of the DFT + BCS
	problem are represented on a three-dimensional Cartesian coordinate
	space. The derivatives of relevant wave functions and densities are
	evaluated using a finite-difference method. For the static problem, the
	Hartree-Fock (HF) nonlinear equations are solved by using the imaginary
	time step method. For the dynamic calculations, the time propagator is
	approximated by Taylor expansion.}  \\
{\em Restrictions: The program suffers from relatively longer running time for 
	the static and dynamic calculations compared to symmetry restricted codes, or those without the full density functional.}  \\
{\em Unusual features: The program includes the full energy density
	functional derived from the Skyrme interaction, including the tensor 
	and the time-odd contributions. This allows the program to compute 
	the odd-$A$ nucleus on the HF level without losing consistency.}   \\ \\

\end{small}

\section{Introduction}
\label{intro}

Nuclear self-consistent mean field models are widely used to describe
ground-state and low-energy properties of medium and heavy
nuclei~\cite{nege82}. The power of these models lies in the fact that the
mean-field potentials are calculated variationally from the effective
nucleon-nucleon interaction. For example, with only about ten parameters, the
Skyrme Hartree-Fock (SHF) model can describe the available experimental nuclear
masses with an error of only $\approx1$\,MeV. The nuclear density-functional
theories (DFTs) are similar to the SHF models, except that the energy density
functionals (EDF) of the DFTs may not correspond to an effective
nucleon-nucleon force.

With the time-dependent DFT (TDDFT), the dynamics of finite nuclei can be
studied~\cite{naka16,sime18,stev19,seki19,tohy20}. These include the harmonic
vibrations near the ground state (g.s.) and the large amplitude motions such as
nuclear fission and fusion.

One of the main advantages of the TDDFT is that these models are in some sense
parameter-free, in that the parameters from the effective interactions are
fitted to the g.s. data and nuclear matter properties, with no further
adjustment to dynamics. However, older TDDFT implementations omitted some of
the terms arising from the Skyrme interaction in order to make the computation
mode tractable. For example, the earliest TDHF calculations for heavy ion
reactions used a simplified Skyrme interaction~\cite{bonche78}. Calculations
omitting the spin-orbit force showed an anomolously low fusion cross section
for central collisions, an effect which was resolved by the inclusion of
spin-orbit \cite{umar_resolution_1986}. Full spin-orbit contributions of the
Skyrme interaction are available in Refs.~\cite{naka05,umar06,maruhn06}, their
inclusion ultimately requiring the increased computing power then available.
The tensor contribution, which plays a role in describing the shell evolutions
near spherical shell closures~\cite{otsuka05,colo07}, has been included in TDHF
calculations~\cite{dai14,stev16,guo18a,guo18b,godbey19}, though no published
codes with these terms are available. See Ref.~\cite{stev19} for a recent
review of the history and role of terms in the density functional and their
application in TDHF/TDDFT.

The purpose of the
current work is to present a package of codes that enable one to perform static
and dynamic simulations on the same footing. For the static calculations, a
large number of computer
codes~\cite{benn05,carl10,stoi05,perez17,doba97a,doba97b,doba04,doba09,schunck12,bonche05}
exist. The code in this work is advantageous in that we do not have limitations
on the spatial coordinates or truncation of terms in the Skyrme EDFs. For the
TDDFT calculations, similar codes~\cite{maruhn14,Schuetrumpf2018,jin21} exist but with
different emphasis on their functionality.

Compared to a previously published code, Sky3D~\cite{maruhn14,Schuetrumpf2018},
that belongs to a similar category, the code \HIT~contains three main features
that differentiate itself. First, it could perform constrained calculations in
the static case.  The static part of the code possesses the same function as
the Ev8 code~\cite{ryss15a}, except that \HIT~relaxes spatial and time-reversal
symmetries. In addition, the \HIT~code contains time-odd densities facilitating
the inclusion of single-particle (s.p.) Hamiltonians of terms that break the
time-reversal symmetry of an even-even nucleus. Second, compared to the Sky3D
code, the \HIT~code includes the tensor interaction in the static and dynamic
calculations. Third, since the \HIT~code is written in C++ language, it can be
parallelized efficiently using C++ Standard Parallelism or C++-based CUDA.

Section~\ref{models} presents the theoretical models and the numerical methods
used in the \HIT~code. In Sec.~\ref{comparisons}, we compare the static and
dynamic results of \HIT~code with those of other codes. In
Sec.~\ref{inputs}, we include a detailed description of the inputs for the
code.

\section{Implemented models}
\label{models}

This section presents the theories implemented and the numerical methods used
to solve the static and dynamic problems. First, we briefly introduce the
nuclear DFT (Sec.~\ref{dft}), which is at the core of the methods that the
\HIT~code solves. Second, we discuss the constraint method used in the
\HIT~code (Sec.~\ref{constraints}). This is followed by a discussion of the
time advancement method used in the code (Sec.~\ref{td}). We end the section
with the detailed numerical method used to solve the static and time-dependent
problems (Sec.~\ref{fd}).

\subsection{The DFT method}
\label{dft}

There have been many reviews (see e.g.~\cite{bender03}) of the nuclear DFT. Here, we focus
on aspects that are relevant to the time-dependent application, as well as the
tensor parts which have not been standardized among theoretical groups.

\subsubsection{The single-particle states}

In a nuclear Hartree-Fock (HF) theory, one describes the many-body system in
terms of an anti-symmetrized product of s.p. wave functions $\phi_k$. The
Bardeen-Cooper-Schrieffer (BCS) theory postulates the following many-body wave
function for an even-even nucleus
\begin{equation}
	\label{BCS}
	|{\rm BCS}\rangle =\prod_{k>0}\big(u_k+v_k\hat{a}_k^+\hat{a}_{\bar{k}}^+)|0\rangle,
\end{equation}
where $|0\rangle$ is the vacuum state, $\hat{a}_k^+$ the generator of a fermion
in s.p. state $\phi_k$, and $\hat{a}_{\bar{k}}^+$ the time reverse partner to
$\phi_k$. The $u_k$ and $v_k$ are linked by the normalization condition
$u_k^2+v_k^2=1$. The phase convention used is $u_k=u_{\bar{k}}\ge0$ and
$v_k=-v_{\bar{k}}\ge0$. That is, the $u_k$'s and $v_k$'s are chosen as positive
real numbers.

The index, $q\in\{n,p\}$, labels neutron and proton quantities. 
For a given nucleonic type, the s.p. states with spin degree of freedom in
the Cartesian coordinate representation can be written in a spinor form 
\begin{equation}
	\phi_{k,q}(\bm{r}\sigma)=\mqty(\phi_{k,q}(\bm{r}\sigma=+1) \\ 
	\phi_{k,q}(\bm{r}\sigma=-1) ),
\end{equation}
where $\phi_{k}(\bm{r}\sigma)$ represents the spin-up and spin-down components
for $\sigma=+1$ and $-1$, respectively.

For an even-even nucleus, the s.p. Hamiltonian for the static g.s. contains no term
that breaks the time-reversal symmetry. Consequently, the resulting s.p. states
are doubly degenerate. In addition, the majority of the application scenarios
involves nuclei that possess certain spatial symmetry. Enforcing the
time-reversal or certain spatial symmetry can significantly reduce the
computing time involved in the orthonormalization step~\cite{ryss16}. However,
in the \HIT~code, we do not enforce these two types of symmetries. That is, the
s.p. wave functions are symmetry-unrestricted 3D quantities.

\subsubsection{The Skyrme Interaction}
\label{skyrme_int}

One of the most popular nuclear DFTs is derived from an effective interaction
proposed by Skyrme~\cite{skyrme58,vaut72}. It is composed of central,
spin-orbit (LS), and tensor interactions 
\begin{equation}
	\label{skyrme_force}
	\hat{v}=\hat{v}^{\rm central}+\hat{v}^{\rm LS}+\hat{v}^{\rm tensor}.
\end{equation}
The two-body Skyrme interaction is defined as 
\begin{equation}\label{skyrme_central}
\begin{aligned}
\hat{v}^{\rm central}(\vb*{r},\vb*{r'}) &= t_0\,\qty(1+x_0\hat{P}_\sigma)\,\delta(\vb*{r}-\vb*{r'}) \\
& \quad +\frac{1}{2}\,t_1\,\qty(1+x_1\hat{P}_\sigma)\,\qty[\hat{\vb*{k'}}^2\,\delta(\vb*{r}-\vb*{r'})+\delta(\vb*{r}-\vb*{r'})\,\hat{\vb*{k}}^2]\\
& \quad +t_2\,\qty(1+x_2\hat{P}_\sigma)\,\hat{\vb*{k'}}\cdot\delta(\vb*{r}-\vb*{r'})\,\hat{\vb*{k}} \\
& \quad +\frac{t_3}{6}\,\qty(1+x_3\hat{P}_\sigma)\,\rho^{\alpha}\qty(\frac{\vb*{r}+\vb*{r'}}{2})\,\delta(\vb*{r}-\vb*{r'}),
\end{aligned}
\end{equation}
where
$\hat{P}_\sigma\equiv\frac{1}{2}\qty(1+\hat{\vb*{\sigma}}_1\cdot\hat{\vb*{\sigma}}_2)$
is the spin exchange operator,
$\hat{\vb*{k}}\equiv\frac{1}{2i}\qty(\vb*{\nabla}_1-\vb*{\nabla}_2)$ is the
relative momentum operator acting to the right, 
$\hat{\vb*{k'}}\equiv-\frac{1}{2i}\qty(\vb*{\nabla}'_1-\vb*{\nabla}'_2)$ acting to the left, 
and $\rho\equiv\rho_n+\rho_p$ is the total density. 
The spin-orbit part of the interaction is given by
\begin{equation}
\label{skyrme_ls}
	\hat{v}^{\rm LS}=i\,W_0\,\qty(\hat{\vb*{\sigma}}_1+\hat{\vb*{\sigma}}_2)\cdot\hat{\vb*{k'}}\times\delta(\vb*{r}-\vb*{r})\,\hat{\vb*{k}}.
\end{equation}
The tensor part of the Skyrme interaction is given by
\begin{equation}\label{skyrme_ten}
\begin{aligned}
\hat{v}^{\rm tensor}&(\vb*{r},\vb*{r'})  \\
	&=\frac{1}{2}t_e\left\{\qty[3\qty(\hat{\vb*{\sigma}}_1\cdot\hat{\vb*{k}})\qty(\hat{\vb*{\sigma}}_2\cdot\hat{\vb*{k'}})-\qty(\hat{\vb*{\sigma}}_1\cdot\hat{\vb*{\sigma}}_2)\hat{\vb*{k'}}^2]\,\delta(\vb*{r}-\vb*{r'})\right.  \\
	&\quad \left.+\delta(\vb*{r}-\vb*{r'})\,\qty[3\qty(\hat{\vb*{\sigma}}_1\cdot\hat{\vb*{k}})\qty(\hat{\vb*{\sigma}}_2\cdot\hat{\vb*{k}})-\qty(\hat{\vb*{\sigma}}_1\cdot\hat{\vb*{\sigma}}_2)\hat{\vb*{k}}^2]\right\}  \\
	&\quad +t_o\left\{\qty[3\qty(\hat{\vb*{\sigma}}_1\cdot\hat{\vb*{k'}})\qty(\hat{\vb*{\sigma}}_2\cdot\hat{\vb*{k}})-\qty(\hat{\vb*{\sigma}}_1\cdot\hat{\vb*{\sigma}}_2)\hat{\vb*{k'}}\cdot\hat{\vb*{k}}]\,\delta(\vb*{r}-\vb*{r'})\right\}.
\end{aligned}
\end{equation}

\subsubsection{Densities in the Skyrme energy density functional}
\label{local_densities}

When performing the HF calculations~\cite{ring80}, one first evaluates the
expectation value of the Hamiltonian over a Slater determinant, which
consists of a set of trial s.p. wave functions. Due to the structure of the
Skyrme force (\ref{skyrme_force}), the resulting total energy turns out to be an
integral of a functional of various local densities~\cite{vaut72}. This
functional, which is derived from the Skyrme force (\ref{skyrme_force}), is
called the Skyrme EDF.

The local densities can be constructed from the non-local 
density, which reads~\cite{doba00b,perl04}
\begin{subequations}
\begin{equation}
	\label{nonlocal_dens1}
	\rho_q(\bm{r}\sigma,\bm{r'}\sigma') = \frac{1}{2}\rho_q(\bm{r},\bm{r'})\delta_{\sigma\sigma'}
					    +\frac{1}{2}\mel{\sigma}{\vu*{\sigma}}{\sigma'}\cdot\bm{s}_q(\bm{r},\bm{r'}),
\end{equation}
with
\begin{align}
        \rho_q(\bm{r},\bm{r'}) &= \sum_{\sigma=\pm1}\rho_q(\bm{r}\sigma,\bm{r'}\sigma),\label{rhoq}\\
	\bm{s}_q(\bm{r},\bm{r'}) &= \sum_{\sigma\sigma'=\pm1}\rho_q(\bm{r}\sigma,\bm{r'}\sigma')\mel{\sigma'}{\vu*{\sigma}}{\sigma},\label{s_q}
\end{align}
\end{subequations}
where $\hat{\sigma}_{x,y,z}$ are the Pauli matrices.

In terms of the s.p. wave functions, the nonlocal density matrix
(\ref{nonlocal_dens1}) is given by
\begin{equation}
	\label{nonlocal_dens2}
	\rho_q(\bm{r}\sigma,\bm{r'}\sigma')=\sum_{k}\,v_{k,q}^2\,\phi_{k,q}(\bm{r}\sigma)\,\phi_{k,q}^*(\bm{r'}\sigma'),
\end{equation}
where the occupation factors $v_{k,q}^2$ are determined through the BCS
procedure (Sec.~\ref{BCS_pro}) based on the s.p. energies obtained at each
iteration. Inserting Eq.~(\ref{nonlocal_dens2}) into Eqs.~(\ref{rhoq}) and
(\ref{s_q}), one obtains the $\rho_q(\bm{r},\bm{r'})$ and
$\bm{s}_q(\bm{r},\bm{r'})$ densities.

With $\rho_q$ (\ref{rhoq}) and $\bm{s}_q$ (\ref{s_q}), the local densities and
currents used to construct the Skyrme energy density functional are given by
\begin{subequations}
	\label{local_densities}
\begin{align}
\rho_q(\bm{r}) &= \rho_q(\bm{r},\bm{r'})\big|_{\bm{r}=\bm{r'}}, \label{rho1}\\
\tau_q(\bm{r}) &= \div{\grad'\,\rho_q(\bm{r},\bm{r'})}\big|_{\bm{r}=\bm{r'}},\\
J_{q,\mu\nu}(\bm{r}) &= \frac{1}{2i}(\nabla_{\mu}-\nabla'_{\mu})\,s_{q,\nu}(\bm{r},\bm{r'})\big|_{\bm{r}=\bm{r'}},\\
\bm{j}_q(\bm{r}) &= \frac{1}{2i}(\grad-\grad')\,\rho_q(\bm{r},\bm{r'})\big|_{\bm{r}=\bm{r'}},\\
\bm{s}_q(\bm{r}) &= \bm{s}_q(\bm{r},\bm{r'})\big|_{\bm{r}=\bm{r'}},\\
\bm{T}_q(\bm{r}) &= \div{\grad'\,\bm{s}_q(\bm{r},\bm{r'})}\big|_{\bm{r}=\bm{r'}},\\
	\label{rho7}
{\it F}_{q,\mu}(\bm{r}) &= \frac{1}{2}\sum_{\nu=x,y,z}(\nabla_{\mu}\nabla'_{\nu}+\nabla'_{\mu}\nabla_{\nu})\,s_{q,\nu}(\bm{r},\bm{r'})\big|_{\bm{r}=\bm{r'}},
\end{align}
\end{subequations}
which are the density $\rho_q(\bm{r})$, the kinetic density $\tau_q(\bm{r})$,
the spin-current (pseudotensor) density $J_{q,\mu\nu}(\bm{r})$, the current
(vector) density $\bm{j}_q(\bm{r})$, the spin (pseudovector) density
$\bm{s}_q(\bm{r})$, the spin-kinetic (pseudovector) density $\bm{T}_q(\bm{r})$,
and the tensor-kinetic (pseudovector) density $\bm{F}_q(\bm{r})$.

From the properties of the density and spin density matrices under time 
reversal~\cite{engel75},
\begin{align}
	\rho_q^T(\bm{r},\bm{r'}) &= \rho_q(\bm{r'},\bm{r}),\\
	\bm{s}_q^T(\bm{r},\bm{r'}) &= -\bm{s}_q(\bm{r'},\bm{r}),
\end{align}
it follows that
\begin{align}
	\rho_q^T(\bm{r})&=\rho_q(\bm{r}),~~ \tau_q^T(\bm{r})=\tau_q(\bm{r}),~~ J_{q,\mu\nu}^T=J_{q,\mu\nu}(\bm{r}), \nonumber \\
	\bm{s}_q^T(\bm{r})&=-\bm{s}_q(\bm{r}),~~ \bm{j}_q^T(\bm{r})=-\bm{j}_q(\bm{r}),~~ \bm{T}_q^T(\bm{r})=-\bm{T}_q(\bm{r}), \\
	\bm{F}_q^T(\bm{r})&=-\bm{F}_q(\bm{r}). \nonumber
\end{align}
It can be seen that $\rho_q(\bm{r})$, $\tau_q(\bm{r})$, and
$J_{q,\mu\nu}(\bm{r})$ are time even, whereas $\bm{s}_q(\bm{r})$,
$\bm{j}_q(\bm{r})$, $\bm{T}_q(\bm{r})$, and $\bm{F}_q(\bm{r})$ are time odd.
When time reversal is a self-consistent symmetry, the time-odd densities
vanish. When time reversal is no longer a symmetry, such as those in the TDDFT
calculations, the time-odd densities start to contribute. 

Finally, the pairing densities read
\begin{equation}
\label{pair_dens}
\tilde{\rho}_q(\bm{r})=\sum_{k,\sigma}\,u_{k,q}\,v_{k,q}\,\big|\phi_{k,q}(\bm{r}\sigma)\big|^2.
\end{equation}

In Eqs.~(\ref{nonlocal_dens2}) and (\ref{pair_dens}), the \HIT~code sums over
all s.p. states, without assuming the fact that they are doubly degenerate.
Here, we add a few words about why the code does not enforce such a symmetry.
The Skyrme interaction (\ref{skyrme_force}) we use conserves the time-reversal
symmetry. If we compute the g.s. of an even-even nucleus, the time-reversal
invariance of the s.p. Hamiltonian (\ref{Ham}) will be maintained throughout
the iteration. This self-consistent symmetry leads to the double-degeneracy of
the s.p. states. For odd-$A$ or odd-odd nuclei, the g.s. and the time
advancement require the breaking of the time-reversal symmetry in both phases
of the calculations. For even-even nuclei, the static calculation may benefit
from enforcing the time-reversal symmetry. But the time advancement requires
the breaking of this symmetry.

\subsubsection{The Skyrme energy density functionals}
\label{SEDF}

In the HF method~\cite{ring80}, one starts with a Slater determinant ($\ket{\Phi}$)
comprised of a set of trial s.p. wave functions ($\ket{\phi_k}$). The expectation
value of the Hamiltonian ($\ev{\hat{H}}{\Phi}$) of a collection of nucleons
interacting through the Skyrme two-body interaction is given
by~\cite{engel75,perl04,lesinski07,hellemans12}
\begin{align}
	\label{Etot}
	E &= E_{\rm kin+c.m.} + E_{\rm Skyrme} + E_{\rm pair} + E_{\rm Coul} \nonumber \\
	  &= \int d^3 r \Bigg\{\mathcal{K}(\bm{r}) + \mathcal{H}^{\rm Skyrme}(\bm{r}) \Bigg.\nonumber \\
	  & \quad \Bigg.+ \sum_{q=p,n}\left[\mathcal{H}^{\rm Skyrme}_q(\bm{r}) + \mathcal{H}^{\rm pair}_q(\bm{r})\right] + \mathcal{H}^{\rm Coul}(\bm{r})\Bigg\},
\end{align}
with
\begin{subequations}
	\label{EDF1}
\begin{align}
\mathcal{H}^{\rm Skyrme}(\bm{r}) &= b_1\rho^2+b_3(\rho\tau-\bm{j}^2)+b_5\rho\Delta\rho+b_7\rho^{2+\alpha} \nonumber\\
		 & \quad +b_9\qty[\rho\div{\bm{J}}+\bm{j}\vdot\qty(\curl{\bm{s}})]+b_{10}\bm{s}^2+b_{12}\rho^{\alpha}\bm{s}^2 \nonumber \\
                 & \quad +b_{14}\qty(\sum_{\mu,\nu=x}^zJ_{\mu\nu}J_{\mu\nu}-\bm{s}\vdot\bm{T}) \nonumber \\
                 & \quad +b_{16}\qty[\qty(\sum_{\mu=x}^zJ_{\mu\mu})^2+\sum_{\mu,\nu=x}^zJ_{\mu\nu}J_{\nu\mu}-2\bm{s}\vdot\bm{F}] \nonumber \\
                 & \quad +b_{18}\bm{s}\vdot\Delta\bm{s}+b_{20}(\div{\bm{s}})^2,
\end{align}
where the densities and currents without subscripts denote the total density such as $\rho=\rho_n+\rho_p$, and
\begin{align}
\mathcal{H}^{\rm Skyrme}_q(\bm{r}) &= b_2\rho_q^2+b_4(\rho_q\tau_q-\bm{j}_q^2)+b_6\rho_q\Delta\rho_q+b_8\rho^{\alpha}\rho_q^2  \nonumber\\
		 & \quad +b_{9q}\qty[\rho_q\div{\bm{J}_q}+\bm{j}_q\vdot\qty(\curl{\bm{s}_q})]+b_{11}\bm{s}_q^2+b_{13}\rho^{\alpha}\bm{s}_q^2  \nonumber \\
		 & \quad +b_{15}\qty(\sum_{\mu,\nu=x}^zJ_{q,\mu\nu}J_{q,\mu\nu}-\bm{s}_q\vdot\bm{T}_q)  \nonumber \\
		 & \quad +b_{17}\qty[\qty(\sum_{\mu=x}^zJ_{q,\mu\mu})^2+\sum_{\mu,\nu=x}^zJ_{q,\mu\nu}J_{q,\nu\mu}-2\bm{s}_q\vdot\bm{F}_q]  \nonumber \\
		 & \quad +b_{19}\bm{s}_q\vdot\Delta\bm{s}_q+b_{21}(\div{\bm{s}_q})^2.
\end{align}
\end{subequations}
In Eq.~(\ref{Etot}), the Coulomb energy for protons and the correction energies
are postponed for the next subsections. Here, we focus on the Skyrme EDF, where
the included tensor contributions vary for different theoretical groups.

The terms of EDFs are derived from the Skyrme force. Hence, there is a fixed
link between the $\{t,x\}$ parameters in Eq.~(\ref{skyrme_force}) and the
$\{b\}$ parameters in Eq.~(\ref{EDF1}). This connection is provided
in~\ref{coefficients} through the $\{a\}$ parameters presented in
Ref.~\cite{bart17}. Another class of parameters is obtained by regrouping the
EDF (\ref{EDF1}) as functions of isoscalar and isovector quantities\footnote{In
Eq.~(\ref{EDF1}), the total energy is grouped in such a way that summation runs
over proton and neutron quantities as indicated by $\sum_{q\in\{n,p\}}$. For
the isoscalar-isovector representation, the total energy is composed of
quantities with index $t=0$ for isoscalar densities ($\rho_{t=0}=\rho_n+\rho_p$
and similarly for the other densities) and $t=1$ for isovector densities
($\rho_{t=1}=\rho_n-\rho_p$ etc.).}. This set of parameters, the $\{C_t\}$
parameters, can be linked to the $\{t,x\}$ parameters as presented in
Ref.~\cite{stev19}.

Kohn and collaborators~\cite{hohenberg64,kohn65} proved that, for the g.s. of
interacting electrons, a density functional exists and can be determined by a
procedure similar to that of a HF method. This liberates one from having to
refer to an interaction. Instead, one may try to determine a density functional
for the g.s. A good starting point is an EDF, which contains the same terms as
those derived from the Skyrme force (\ref{EDF1}). That is, the $\{b\}$
parameters are now not necessarily determined from the $\{t,x\}$ parameters. We
can treat those $\{b\}$ parameters as free ones. Having the total energy, we
solve the Kohn-Sham/HF equations to determine the g.s. energy and s.p. wave
functions. This method is what we call the nuclear DFT.

Still, the majority of existing parameters are derived from the Skyrme force.
Only a few new EDFs~\cite{kort10,kort12,kort14} are fitted directly by varying
the $\{b\}$ or $\{C\}$ parameters to match the properties of the finite nuclei
and nuclear matter without referring to the Skyrme force. It has to be
mentioned that, in the \HIT~code, we always ignore the terms with
$b_{18,19,20,21}$. The inclusion of them results in instabilities in the static
calculations~\cite{hellemans12}.

\subsubsection{The kinetic energy and center-of-mass correction}
\label{kinetic}

The kinetic energy is given by
\begin{equation}
	\mathcal{K}(\bm{r})=\frac{\hbar^2}{2m}\,\tau(\bm{r}),
\end{equation}
where $m=\frac{1}{2}(m_n+m_p)$ and $\tau=\tau_n+\tau_p$.
The mean-field method uses a set of localized s.p. states, $\phi_{k,q}(\bm{r},\sigma)$ 
to approximate the nuclear g.s. Consequently, the translational invariance
of the Hamiltonian has been broken. Conventionally, the one-body contribtion due to
the broken translational invariance can be approximately included into 
the kinetic energy through
\begin{equation}
	\mathcal{K}(\bm{r})=\frac{\hbar^2}{2m}\,\tau(\bm{r})\,\qty(1-\frac{1}{A}),
\end{equation}
where $A$ is the mass number. 
The additional coefficient is due to the so-called one-body center-of-mass 
correction, which is included in the \HIT~code.

\subsubsection{The pairing energy}

Another important contribution to the total energy (\ref{Etot}) is the 
pairing energy density, which is given by
\begin{equation}
	\mathcal{H}^{\rm pair}_q(\bm{r})=\frac{1}{4}\tilde{\rho}^2_q(\bm{r})\,G(\bm{r}),
\end{equation}
with
\begin{equation}
	\label{pairing_strengths}
G_q(\bm{r})=\left\{
\begin{array}{ll}
	V_{0,q}                                 & \mbox{delta force}\\
	V_{0,q}\qty[1-\frac{\rho(\bm{r})}{2\rho_0}] & \mbox{mixed force} 
\end{array}\right.\, ,
\end{equation}
where $\rho_0=0.16$\,fm$^{-3}$ is the saturation density of symmetric nuclear matter.

\subsubsection{The single-particle Hamiltonian}
\label{Ham}

The minimization of $E$ (\ref{Etot}) with a constraint to ensure normality of each 
s.p. wave functions,
\begin{equation}
	\label{variation}
	\delta\left(E-\sum_{k,\sigma}\sum_{q\in\{n,p\}}\,\epsilon_{k,q}\,\int\,d^3r\,
	\phi^{\ast}_{k,q}(\bm{r}\sigma)\,\phi_{k,q}(\bm{r}\sigma)\right)=0,
\end{equation}
leads to the HF (or Skyrme-Kohn-Sham) equations~\cite{bart17}:
\begin{equation}
	\label{HF_equation}
	\hat{h}_q\,\phi_{k,q}(\bm{r},\sigma)=\epsilon_{k,q}\,\phi_{k,q}(\bm{r},\sigma),
\end{equation}
with the s.p. Hamiltonian for protons and neutrons given by
\begin{align}
\label{hamiltonian}
\hat{h}_q &= -\div{B_q(\bm{r})\grad}+U_q(\bm{r})+\bm{S}_q(\bm{r})\vdot\hat{\bm{\sigma}} \nonumber \\
&\quad +\frac{1}{2i}\sum_{\mu,\nu=x}^z\qty[W_{q,\mu\nu}(\bm{r})\nabla_{\mu}\hat{\sigma}_{\nu}+\nabla_{\mu}\hat{\sigma}_{\nu}W_{q,\mu\nu}(\bm{r})] \nonumber \\
&\quad +\frac{1}{2i}\qty[\bm{A}_q(\bm{r})\vdot\grad + \div{\bm{A}_q(\bm{r})}] \nonumber \\
&\quad -\grad\vdot\qty[\hat{\bm{\sigma}}\vdot\bm{C}_q(\bm{r})]\grad - \grad\vdot\bm{D}_q(\bm{r}) \hat{\bm{\sigma}} \vdot\grad,
\end{align}
and where the various potentials read
\begin{subequations}
	\label{potentials_full}
	\begin{align}
		W_{q,\mu\nu}(\bm{r}) =& -\sum_{\kappa=x}^z\epsilon_{\kappa\mu\nu}\qty(b_9\nabla_{\kappa}\rho+b_{9q}\nabla_{\kappa}\rho_q) + 2b_{14}J_{\mu\nu} \nonumber \\
			   &+ 2b_{15}J_{q,\mu\nu} + 2b_{16}\qty(J_{\nu\mu}+\sum_{\kappa=x}^z J_{\kappa\kappa}\delta_{\mu\nu}) \nonumber \\
			   &+ 2b_{17}\qty(J_{q,\nu\mu}+\sum_{\kappa=x}^z J_{q,\kappa\kappa}\delta_{\mu\nu})\, , \\
		A_{q,\mu}(\bm{r}) =&-2b_3j_{\mu}-2b_4j_{q,\mu}+b_9(\bm{\nabla}\times\bm{s})_{\mu}+b_{9q}(\bm{\nabla}\times\bm{s}_q)_{\mu}\, , \\
		S_{q,\mu}(\bm{r}) =& - \qty(b_9\curl{\bm{j}}+b_{9q}\curl{\bm{j}_q})_{\mu} + 2b_{10}s_{\mu} + 2b_{11}s_{q,\mu} \nonumber \\
			   &+ 2b_{12}\rho^{\alpha}s_{\mu}+2b_{13}\rho^{\alpha}s_{q,\mu}-b_{14}T_{\mu}-b_{15}T_{q,\mu} \nonumber \\
			   &- 2b_{16}F_{\mu} - 2b_{17}F_{q,\mu} + 2b_{18}\Delta s_{\mu}+2b_{19}\Delta s_{q,\mu} \nonumber \\
			   &- 2b_{20}\nabla_{\mu}(\div{\bm{s}}) - 2b_{21}\nabla_{\mu}(\div{\bm{s}_q})\, ,\\
		C_{q,\mu}(\bm{r})    =& - b_{14}s_{\mu} - b_{15}s_{q,\mu}\, ,\\
		D_{q,\mu}(\bm{r})    =& - 2b_{16}s_{\mu} - 2b_{17}s_{q,\mu}\, .
	\end{align}
\end{subequations}
For protons, one needs to add Coulomb potentials [Eqs.~(20) and (24) in
Ref.~\cite{shi18}]. The detailed expression of $U_q$ can be found in Eq. (18)
of Ref.~\cite{shi18}.  

The potential $B_q(\bm{r})$ containing the effective mass through 
$B_q(\bm{r})\equiv\frac{\hbar^2}{2m^*(\bm{r})}$ is defined as
\begin{equation}
	\label{com}
	B_q(\bm{r})=\frac{\hbar^2}{2m}\qty(1-\frac{1}{A})+b_3\rho(\bm{r})+b_4\rho_q(\bm{r}).
\end{equation}

\subsubsection{The Bardeen-Cooper-Schrieffer method}
\label{BCS_pro}

At each iteration for solving the nonlinear Kohn-Sham/HF equation
(\ref{HF_equation}), one obtains the Lagrangian multipliers, $\epsilon_{k,q}$,
which are identified with s.p. energies. We use the BCS
method~\cite{bardeen57} to approach the nuclear pairing interaction.
For BCS method, we associate with each s.p. wave function a real number, $v_k$, whose
square gives the occupation probability of the $k$th orbit.


After each HF iteration, the occupation amplitude $v_k$ is determined, in the
current work, by the following BCS equations~\cite{bender00}
\begin{equation}
\label{occupation}
v_{k,q}^2=\frac{1}{2}\left[1-\frac{\epsilon_{k,q}-\lambda_{q}}{\sqrt{(\epsilon_{k,q}-\lambda_q)^2+\Delta_{k,q}^2}}\right],
\end{equation}
where $\epsilon_{k,q}$'s are the HF s.p. energies; $\lambda_q$ is the
Fermi energy for given nucleonic type, which is adjusted so that $2\sum_k v_{k,q}^2$
gives the correct nucleon number. In Eq.~(\ref{occupation}), the
state-dependent s.p. pairing gaps, $\Delta_{k,q}$'s, are given by
\begin{equation}
\Delta_{k,q} = \sum_{\sigma}\,\int d\bm{r}\,\Delta_q(\bm{r})\,\phi_{k,q}^*(\bm{r},\sigma)\,\phi_{k,q}(\bm{r},\sigma),
\end{equation}
where
\begin{equation}
\label{densities_bcs}
	\Delta_q(\bm{r}) = -\frac{1}{2}\,G_q(\bm{r})\,\tilde{\rho}_q(\bm{r}).
\end{equation}

\subsection{Deformation constraints}
\label{constraints}

In many nuclear mean-field calculations, one needs to calculate the total
energy of the nucleus at a specific deformation. For example, the
generator-coordinate methods use deformation as a generator coordinate to
include more correlations into the g.s. Spontaneous fission calculations
may require potential-energy-surface calculations with high-order deformations. For
some particular configurations of a nucleus, such as the shape isomeric states,
it is favorable to constrain the deformation to a nearby deformation before
removing the constraints, facilitating the variation procedure to converge to
the desired isomeric state.

In TDDFT calculations~\cite{maruhn14,Schuetrumpf2018}, as a starting point,
the center of mass may drift from the center of the simulating box. A deformed
heavy nucleus may rotate with respect to the principal system, resulting in a
slightly rotated static solution before time development. If the initial
guessed wave functions are obtained with the Hamiltonian containing relevant
symmetries, then the self-consistent process tends to maintain the center of
mass at the center of the box or the orientation of the principal system to
align with the laboratory system~\cite{shi20}.

However, for some deformed calculations, it has been noticed that the
self-consistent symmetry could not ensure the nucleus aligns with the
laboratory system~\cite{maruhn14,Schuetrumpf2018}. Hence, it is necessary to be
able to fix the orientation and center of mass of the nucleus in the static
calculation by constraining relevant moments. The \HIT~code allows for precise
constraints on a few multipole moments. Section~\ref{moments} describes the
multipole moments used in the \HIT~code.  In Sec.~\ref{augmented}, we introduce
the method of constraining, namely, the augmented Lagrangian method
(ALM)~\cite{stas10,ryss15a}.

\subsubsection{Multipole moments considered}
\label{moments}

The \HIT~code constrains on two types of nuclear isoscalar multipole moments.
The first type of moments are those that characterize the nucleus' center of
mass and orientations in a simulating box
\begin{align}
	\label{qxyz}
	&\hat{Q}_x\equiv\hat{x},~~\hat{Q}_y\equiv\hat{y},~~\hat{Q}_z\equiv\hat{z}, \\
	\label{qxy}
	&\hat{Q}_{xy}\equiv\hat{x}\hat{y},~~\hat{Q}_{yz}\equiv\hat{y}\hat{z},~~\hat{Q}_{xz}\equiv\hat{x}\hat{z}.
\end{align}
The second type concerns the intrinsic deformations of a nucleus. 
The quadrupole and octupole moments, whose operators are defined as
\begin{align}
	\label{q20_q22}
	&\hat{Q}_{20}\equiv2\hat{z}^2-\hat{x}^2-\hat{y}^2,~~\hat{Q}_{22}\equiv\sqrt{3}(\hat{x}^2-\hat{y}^2), \\
	\label{q30}
	&\hat{Q}_{3\mu}\equiv \Re\qty[\hat{r}^3\hat{Y}^*_{3\mu}(\theta,\phi)],~~\mu=0,1,2,3,
\end{align}
where the spherical harmonics defined in Eq.~(\ref{q30}) can be found in
Ref.~\cite{jackson}. Conventionally, the definitions of $\hat{Q}_{20}$ and $\hat{Q}_{22}$
[Eq.~(\ref{q20_q22})] differ from a spherical harmonics definition.
To make the discussions in later sections more compact, 
we define a new set of multipole moment operators
\begin{equation}
	\label{quadrupole_sh}
	\hat{\mathscr{Q}}_{\lambda\mu}\equiv \Re\qty[\hat{r}^\lambda\hat{Y}^*_{\lambda\mu}(\theta,\phi)].
\end{equation}
For the quadrupole moment operators, the two definitions of
Eqs.~(\ref{q20_q22}) and (\ref{quadrupole_sh}) differ by the coefficients. The
two definitions are the same for the octupole moment operators. For the
static constraint calculations, we use the quadrupole moment operators defined in
Eq.~(\ref{q20_q22}). For time-dependent harmonic vibration
calculations in Sec.~\ref{td}, we use the operator $\hat{\mathscr{Q}}_{\lambda\mu}$
presented in Eq.~(\ref{quadrupole_sh}).

In a static calculation of a triaxial nucleus, the nucleus is placed at the
center of the simulating box. It is oriented in such a way that the principal
axes of the nucleus coincide with the $x$-, $y$-, and $z$-axis of the
laboratory system. These two goals are achieved by constraining the expectation
values of the first type of operators listed in Eqs.~(\ref{qxyz}) and
(\ref{qxy}) to zero.

\begin{figure}[tbp]
\centering
\includegraphics[scale=0.10]{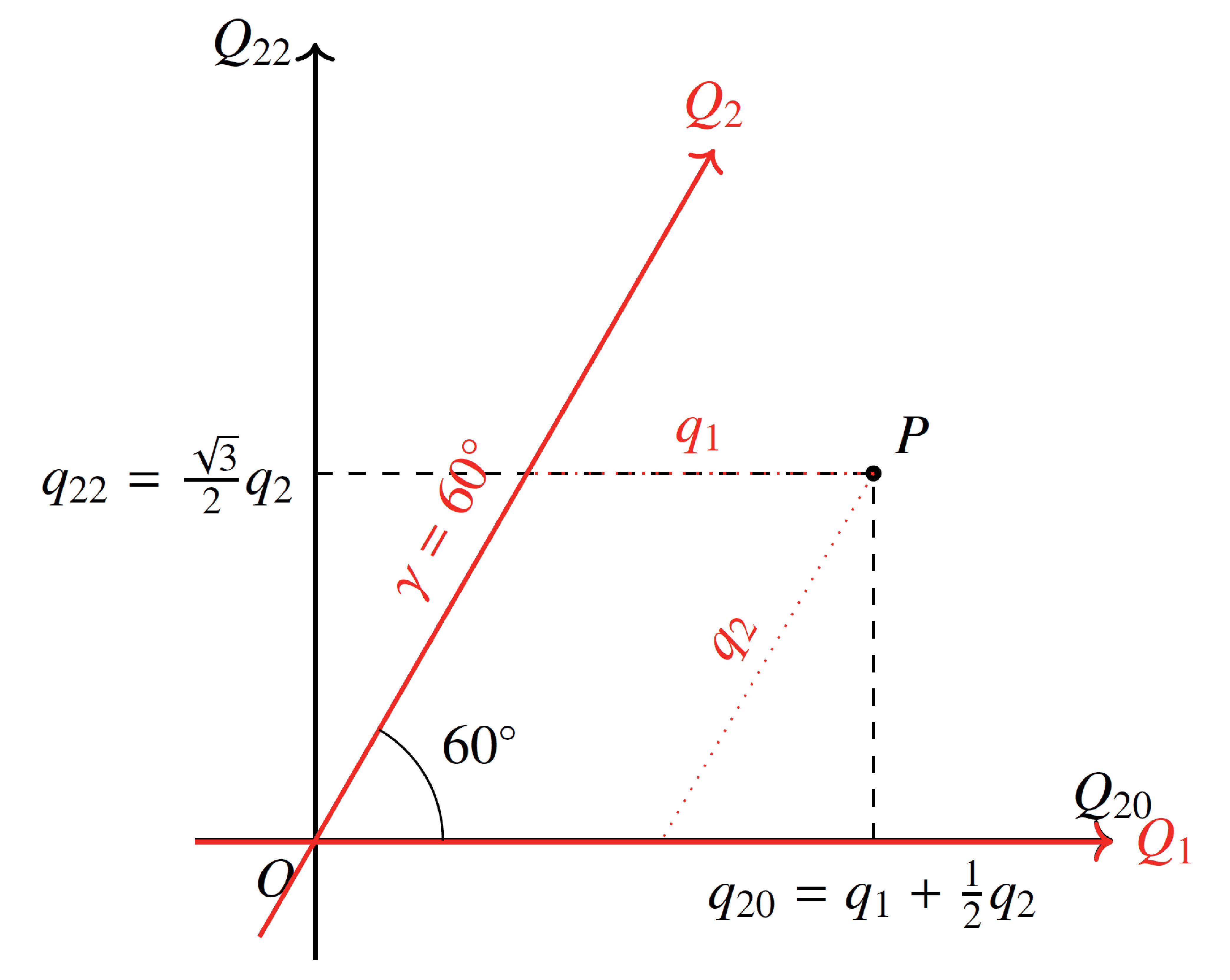}
	\caption{Schematic plot demonstrating the fact that the relation
	between $q_{20,22}$ and $q_{1,2}$~(\ref{q1_q2}) indicates a tilted axis
	for $q_1$ and $q_2$. 
	A quadrupole deformation point ``P'' is at ($q_{20}$, $q_{22}$),
	with $Q_{20}$ and $Q_{22}$ being the horizontal and vertical
	axes ($Q_{20}$-$Q_{22}$ system), respectively. 
	The same point, ``P'' can be represented in a new system formed by $Q_1$ 
	and $Q_2$ (differentiated from the $Q_{20}$-$Q_{22}$ system by red color).  
	In this new system, the coordinate ($q_1$, $q_2$) relates to the
	coordinate in the $Q_{20}$-$Q_{22}$ system through Eq.~(\ref{q1_q2}).}
\label{fig_q1q2}
\end{figure}

Conventionally, one adds constraints on $q_{20}\equiv\langle\hat{Q}_{20}\rangle$ 
and $q_{22}\equiv\langle\hat{Q}_{22}\rangle$ moments for quadrupole deformations. 
However, the \HIT~code adds constraints to the $q_1$ and $q_2$ values. 
The $q_1$ and $q_2$ moments are defined to be linked to the
$q_{20}$ and $q_{22}$ values through~\cite{ryss15a}
\begin{equation}
\label{q1_q2}
\begin{aligned}
q_{20} &= q_1+\frac{1}{2}q_2, \\
q_{22} &= \frac{\sqrt{3}}{2} q_2.
\end{aligned}
\end{equation}
It is customary to define two mass-independent parameters to quantify
the degree of quadrupole deformation, namely
\begin{align}
	 \beta_2&= \sqrt{\alpha_{20}^2+\alpha_{22}^2}, \\
	 \gamma &= 2\arctan(\frac{\sqrt{3}q_2}{\sqrt{(2q_1+q_2)^2+3q_2^2}+2q_1+q_2}),
\end{align}
with $\alpha_{2\mu}=\frac{\sqrt{5\pi}}{3AR^2}q_{2\mu}$, $R=1.2A^{1/3}$\,fm.

From Eq.~(\ref{q1_q2}), it is clear that if $q_{20}$ and $q_{22}$ form a
coordinate system measuring the $x$ and $y$ directions, $q_1$ and $q_2$ form a
system with $q_1$ measuring the horizontal direction, $q_2$ measuring the
direction that is rotated anti-clockwise by $\gamma=60^{\circ}$ with respect
to $q_1$ direction. For details, see Figure~\ref{fig_q1q2}.

\begin{table}[htb]
	\centering
	\caption{Suppose $a$ is a positive real number, the $q_1$ and $q_2$ values
	one needs to take to obtain the typical $\gamma$ values.}
\label{tab_q1_q2}
\begin{tabular}{ccc}
\hline
\hline
	$\gamma$ &  $q_1$ & $q_2$  \\
       \hline
	$+60^{\circ}$ & 0  & $a$  \\
	$+30^{\circ}$ & $a$  & $a$  \\
	$0^{\circ}$ & $a$  & 0    \\
	$-30^{\circ}$ & $2a$  & $-a$  \\
	$-60^{\circ}$ & $a$  & $-a$  \\
	$-90^{\circ}$ & $a$  & $-2a$  \\
	$-120^{\circ}$ & 0  & $-a$  \\
\hline
\hline
\end{tabular}
\end{table}

Compared to the $q_{20}$ and $q_{22}$ values, constraining on the $q_1$ and
$q_2$ values provides more convenient access to the deformation points on the
$\gamma=30^{\circ}$ and 60$^{\circ}$ lines. Indeed, to have a
$\gamma=30^{\circ}$ or 60$^{\circ}$ deformation, one constrains $q_1=q_2={\rm
constant}$ or $q_1=0$, $q_2={\rm constant}$. For applications breaking the
time-reversal symmetry, the deformation points in the sector from
$\gamma=0^{\circ}$ to 60$^{\circ}$ are not sufficient. For example, a cranking
calculation in the $y$ direction may require the calculation of total Routhian
in the range of $\gamma$ from $-120^{\circ}$ to $60^{\circ}$.
Table~\ref{tab_q1_q2} lists a few typical $\gamma$ values and $(q_1,q_2)$
values corresponding to them.

To obtain a potential-energy curve with axial symmetry, one asks for $q_2=0$
and varies $q_1$ from a negative value to a positive value, which corresponds
to a shape varying from an oblate to a prolate deformation. An alternative way
to access the oblate deformation is to keep $q_1=0$ and increase the $q_2$
value to positive numbers. These two methods differ in that the oblate shapes
have different orientations in the Cartesian coordinate. To have a triaxial
deformation, one needs to constrain the $q_2$ to a non-zero value.

\subsubsection{The augmented Lagrangian method}
\label{augmented}

In a quadratic penalty method, one minimizes the following total energy
\begin{equation}
	\label{quad}
	E'=E+C\qty(q-q^0)^2,
\end{equation}
where $q\equiv\langle \hat{Q} \rangle$ and $q^0$ is the desired 
multipole moment introduced in Section~\ref{moments}. 
For each different multipole constraint, a term such as that in
Eq.~(\ref{quad}) is added, and in general, $q$, $q^0$, and the associated
coefficient $C$ have indices to label the particular moments constrained.
If one adds more than one constraint, a summation of terms is assumed. The
indexes and summation symbols are ignored here for brevity.

Taking the variational derivative of $E'$, one obtains
the s.p. Hamiltonian
\begin{equation}
	\label{quad_pot}
	\hat{h'}=\hat{h}+2C\qty(q-q^0)\hat{Q},
\end{equation}
where the parameter $C$ is chosen in such a way that the total constraint
energy $E_{\rm mult}\equiv C(q-q^0)^2$ is in the order of a few MeV at the
beginning of the iteration.

In general, the quadratic penalty method can constrain the solution to be near
the desired value $q^0$, but seldomly to $q^0$ exactly. The ALM is introduced 
to the DFT in Ref.~\cite{stas10}. In the ALM, 
one minimizes the following total energy
\begin{equation}
	\label{alm}
	E'=E+L\qty(q-q^0)+C\qty(q-q^0)^2.
\end{equation}
Compared to the quadratic penalty method in Eq.~(\ref{quad}), the ALM includes 
a linear term in the total energy in Eq.~(\ref{alm}).  The s.p. Hamiltonian becomes
\begin{equation}
	\label{alm_pot}
	\hat{h'}=\hat{h}+2C\qty(q-q^0(L)),
\end{equation}
where $q^0(L)=q^0-\frac{L}{2C}$. Comparing Eq.~(\ref{quad_pot}) and
Eq.~(\ref{alm_pot}), we see that the ALM requires the
targeted multipole moment ($q^0$) to be substituted by $q^0-\frac{L}{2C}$.
However, the $L$ value needs to be updated during the iteration process according to
\begin{equation}
	L^{(k+1)}=L^{(k)}+2C\qty(q-q^0).
\end{equation}
The iterations can start from a zero value, $L^{(0)}=0$.

Another similar constraining method to the ALM has been used in the Lyon 
group~\cite{ryss15a}. The two methods differ in that the targeted
moment is updated according to
\begin{equation}
	\label{L_update}
	L^{(k+1)}=L^{(k)}-\epsilon_{\rm constr.}\qty(q-q^0),
\end{equation}
where $\epsilon_{\rm constr.}$ is a constant between 0 and 1. The two methods
turn out to be effective in constraining the nuclear deformations exactly
at the desired moment values.

\subsection{TDDFT method}
\label{td}

This section presents aspects related to the calculations of nuclear dynamics.
Since the application areas of the TDDFT method are mostly nuclear vibrations
or collisions, the preparations of the initial states in Sec.~\ref{boosts} are
for these two applications. The time evolution of the wave functions is
discussed in Sec.~\ref{evolutions}. Section~\ref{consistencies} includes
comments about the consistencies with the EDFs and pairing treatment of the
static calculations. After the time evolution, analysis needs to be performed.
For the vibration calculations, we include a discussion about the absorbing
boundary conditions in Sec.~\ref{abc}, the Fourier transform of the response
functions in Sec.~\ref{fourier}, and the calculation of the energy-weighted sum
rules in Sec.~\ref{ewsr}.

\subsubsection{Boosting the wave functions}
\label{boosts}

A time-dependent simulation starts with a boost on the static states before
time evolution. Two main research areas exist for the TDDFT application,
namely, the harmonic vibration of nuclear near the minimum of the potential
well and the heavy-ion collision of two nuclei. In this subsection, we discuss
the two ways of exciting the two types of modes in the \HIT~code.

{\it A. Harmonic vibration}

One of the main application areas of the TDDFT is to simulate the harmonic
vibrations of a nucleus and analyze the spectrum. One of the most common
excitation modes is the isovector dipole resonances. To initiate such a
vibration, one applies an instantaneous boost on proton and neutron
wave functions in opposite directions. For details, see Ref.~\cite{shi20}.

This section discusses how the \HIT~code excites multipole vibration modes.
Taking the isoscalar quadrupole boosts as an example, we have
\begin{equation}
\label{boost_isoscalar}
	\phi_{k,q}(\bm{r},\sigma;t=0)=\exp\left[-i\epsilon\,\alpha\,\hat{\mathscr{Q}}_{2,\mu}\right]
	\phi^{\rm(stat)}_{k,q}(\bm{r},\sigma),
\end{equation}
where the constant, $\alpha$, is chosen to be $Z/A$.
Similar choice is made in Refs.~\cite{stoi11,hino19}.
The s.p. wave functions resulted from a converged static
calculation are denoted with $\phi^{\rm(stat)}$. 
For the isovector quadrupole boosts, we differentiate protons and neutrons,
namely
\begin{equation}
\label{boost_isovector}
	\phi_{k,q}(\bm{r},\sigma;t=0)=\exp\left[-i\epsilon\,\alpha_q\,\hat{\mathscr{Q}}_{2,\mu}\right]
	\phi^{\rm(stat)}_{k,q}(\bm{r},\sigma),
\end{equation}
where $\alpha_n=-Ze/A$ and $\alpha_p=Ne/A$.

After the boosts, the quadrupole moments $q^{\rm IS}_{2\mu}(t) \equiv
\ev*{\alpha\,\hat{\mathscr{Q}}_{2,\mu}}{\Phi(t)}$ and $q^{\rm IV}_{2\mu}(t) \equiv
\sum_{q\in\{n,p\}} \ev*{\alpha_q\,\hat{\mathscr{Q}}_{2,\mu}}{\Phi(t)}$ are
recorded for the isoscalar and isovector boosts, respectively. These quadrupole
moments are later used to perform the Fourier analysis to obtain the strength
functions. The current version of the \HIT~code allows for the isoscalar and
isovector boosts with $(\lambda,\mu)=(2,0), (2,1), (2,2), (3,0), (3,2)$. In
addition, it boosts isoscalar monopole [$(\lambda,\mu)=(0,0)$] and isovector
dipole ($\lambda=1$) moments.

{\it B. Heavy-ion reactions}

To prepare for reaction calculations involving two nuclei, one first
places the nuclei at the proper positions before giving them momenta. This is done
by the following transformation~\cite{maruhn14,stev19}
\begin{subequations}
	\label{boost_trans}
	\begin{equation}
	\phi_{k,q}^{(1)}(\bm{r},\sigma;t=0)=\exp(i \bm{p}_1\cdot\bm{r})\,\phi^{\rm (stat)}_{k,q}(\bm{r}-\bm{R}_1,\sigma),
	\end{equation}
	\begin{equation}
	\phi_{k,q}^{(2)}(\bm{r},\sigma;t=0)=\exp(i \bm{p}_2\cdot\bm{r})\,\phi^{\rm (stat)}_{k,q}(\bm{r}-\bm{R}_2,\sigma).
	\end{equation}
\end{subequations}
In this case, two identical nuclei, denoted with $(1)$ and $(2)$, are placed at
$\bm{R}_1$ and $\bm{R}_2$, respectively. They are boosted from a g.s.
indicated by $\phi^{\rm (stat)}$. The resulting two nuclei move with the momenta of
$\bm{P}_1=A_1\bm{p}_1$ and $\bm{P}_2=A_2\bm{p}_2$.

Note, that since the wave functions are defined only on the grid points, we can
{\it only} move nucleus at incremented positions. To place the nuclei at any
position, one needs to perform an interpolation procedure. This may end up with
an excited HF solution. Hence, the current version of the \HIT~code does not
have this feature of repositioning the nucleus at any coordinate in the
simulating box.

Upon the boosts in Eqs.~(\ref{boost_isoscalar}) and (\ref{boost_trans}) being 
performed, a new compound system is formed. For the nuclear reaction case, the
system constitutes two nuclei placed at a distance:
\begin{align*}
	\phi_{k,q}(\bm{r},\sigma)&=\phi_{k,q}^{(1)}(\bm{r},\sigma), ~~ \mbox{for } k=1,...,M, \nonumber \\
	\phi_{M+k,q}(\bm{r},\sigma)&=\phi_{k,q}^{(2)}(\bm{r},\sigma), ~~ \mbox{for } k=1,...,N. \nonumber
\end{align*}
The new compound system $\phi(\bm{r},\bm{\sigma})$ is formed by concatenating the s.p. wave function indices of the 
two displaced systems (\ref{boost_trans}) together. Before
performing time advancement, the code performs an orthonormalization for this
new system.

\subsubsection{The time advancement}
\label{evolutions}

The nuclear non-relativistic time-dependent Schr\"odinger equation reads
\begin{equation}
	\label{tdhf}
	i\hbar\pdv{\phi_{k,q}(t)}{t}=\hat{h}_q(t)\phi_{k,q}(t),
\end{equation}
where $\hat{h}_q$ can be found in Eq.~(\ref{hamiltonian}). In this section, the
subscript $q$ is ignored for simplicity. The equation (\ref{tdhf}) has the
formal solution
\begin{equation}
	\label{evolve}
	\phi_k(t)=\hat{\mathscr{U}}(t)\phi_k(0)=\hat{T}\exp(-\frac{i}{\hbar}\int_0^t\hat{h}(t')\,dt')\phi_k(0),
\end{equation}
where $\hat{\mathscr{U}}$ is the time-evolution operator, and $\hat{T}$ is the
time-ordering operator. To solve the time-dependent problem, one breaks up the
total time evolution into $N$ small increments of time $\Delta t$
\begin{equation}
	\hat{U}(t,t+\Delta t)=\exp(-\frac{i}{\hbar}\int_t^{t+\Delta t}\hat{h}(t')\,dt').
\end{equation}

The time-evolution operator $\hat{\mathscr{U}}(t)$ can be obtained by
consecutive actions of $\hat{U}(t,t+\Delta t)$
\begin{equation}
	\hat{\mathscr{U}}(t)=\prod_{n=0}^{N-1}\hat{U}(n\Delta t,(n+1)\Delta t).
\end{equation}

For small $\Delta t$ one could approximate $\hat{U}(t,t+\Delta t)$ by Taylor
expansion up to order $m$:
\begin{equation}
\label{expansion}
	\exp(-\frac{i}{\hbar}\hat{h}\Delta t)\approx\sum_{n=0}^m\frac{1}{n!}\left(\frac{-i\Delta t}{\hbar}\right)^n \hat{h}^n,
\end{equation}
where $\hat{h}$ has been assumed to be time independent in the time interval of
$\Delta t$. In the current work, $\Delta t$ is taken to be 0.2\,fm/$c$, and
$m=4$. These choices are motivated by previous TDHF calculations.

In realistic calculations, each time advance of s.p. wave functions
$\phi_k$, from time $t$ to $t+\Delta t$, has been achieved by using the
procedure adopted by the Sky3D code~\cite{maruhn14,Schuetrumpf2018}. Specifically, 
from a series of s.p. wave functions at $t$, $\phi_k(t)$, one first performs
\begin{equation}
\label{time_progress}
\phi_k^{\rm temp}(t+\Delta t)=\hat{U}^{\rm t}(t,t+\Delta t) \phi_k(t). 
\end{equation}
Having $\phi_k^{\rm temp}(t+\Delta t)$, and $\phi_k(t)$, one assembles various
densities using respective s.p. wave functions, obtaining the
$\rho^{\rm temp}(t+\Delta t)$ and $\rho(t)$.

Using these densities, one obtains the densities at a ``middle time'',
$\rho^{\rm mid}(t+\frac{\Delta t}{2})=0.5[\rho^{\rm temp}(t+\Delta
t)+\rho(t)]$. Now, one constructs the Hamiltonian $\hat{h}^{\rm mid}$, using
$\rho^{\rm mid}(t+\frac{\Delta t}{2})$ [see Eq.~(15) of Ref.~\cite{shi18}, and
Eq.~(\ref{hamiltonian}) for the form of the Hamiltonian]. A second time
propagation operation $\hat{U}^{\rm mid}(t,t+\Delta t)$ with $\hat{h}^{\rm
mid}$ [Eq.~(\ref{expansion})] is performed on the s.p. states,
finally obtaining the wave functions at $t+\Delta t$ 
\begin{equation}
\phi_k(t+\Delta t)=\hat{U}^{\rm mid}(t,t+\Delta t)\phi_k(t).  
\end{equation}
Here, $\hat{U}^{\rm mid}$ differs from $\hat{U}^{\rm t}$
[Eq.~(\ref{time_progress})] in that the former uses the s.p.
Hamiltonian in its exponent [Eq.~(\ref{expansion})] at the time $t+\frac{\Delta
t}{2}$, whereas the latter refers to the operator $\hat{U}$, where the
Hamiltonian is constructed using the quantities at the time $t$.

\subsubsection{Consistencies in the static and dynamic calculations}
\label{consistencies}

As alluded to in the Introduction (\ref{intro}), the \HIT~code ensures that the
EDFs in static and time-dependent calculations are the same. For the kinetic
term (\ref{com}) in $\hat{h}_q$ (\ref{hamiltonian}), the \HIT~code switches on
or off the center-of-mass correction for static and time-dependent parts
together.  For the case of harmonic vibrations, careful analyses~\cite{frac12}
show that small spurious strength may occur if the consistency is not
maintained. In the results presented later in this work, we ignore the
center-of-mass correction for both static and dynamic parts. We also note that
for collisions where the mass number $A$ is different for the fragments and the
combined system, the simple centre-of-mass prescription (\ref{com}) cannot be
used since there is not a single consistent definition of $A$.

When the BCS pairing is included, the occupation amplitudes, $v_{k,q}$'s, are
kept unchanged during the time development~\cite{maruhn14,Schuetrumpf2018}. When
evaluating the densities, the s.p. wave functions vary according to
Eq.~(\ref{tdhf}). This is a coarse approximation of dynamical pairing, as the
occupation probabilities should vary with time. Indeed, some of the problems
associated with the TDHF + BCS method in describing particle transport have
been discussed in Ref.~\cite{scamps12}. This approximation of the pairing will
be improved in our future developements. A natural extension would be to solve
the full time-dependent HFB problem~\cite{stetcu11,bulgac16,magi17}. Since the
HFB theory treats nuclear interactions in the particle-hole and pairing
channels in one single variational process~\cite{ring80}, a time-dependent HFB
treatment allows for the occupation amplitudes to be determined dynamically by
the upper and lower components at a given time.

\subsubsection{The absorbing boundary conditions}
\label{abc}

With Dirichlet boundary conditions, it is known that the TDDFT
calculations show the occurrence of non-physical particle densities at the
boundary region. To cure this problem, the so-called absorbing boundary
conditions (ABC) is proposed~\cite{naka05}. The \HIT~code implements such a
feature by introducing an imaginary potential
\begin{subequations}
\label{eq:ABC}
\begin{equation}
\hat{h}(\bm{r})\rightarrow \hat{h}(\bm{r})+i\tilde{\eta}(\bm{r})
\end{equation}
at the boundary region of the form
\begin{equation}
\tilde{\eta}(\bm{r})=\left\{\begin{array}{ll}
0 & \mbox{for }0<|\bm{r}| \le R\\
\eta_0\frac{|\bm{r}|-R}{\Delta r} & \mbox{for } R<|\bm{r}|<R+\Delta r
\end{array}\right.\, .
\end{equation}
\end{subequations}
The ABC method turns out to be effective in absorbing the emitted particles.

\subsubsection{The Fourier transform}
\label{fourier}

For harmonic vibrations, after the boost, the nucleus starts to vibrate. 
The relevant multipole moment $q(t)\equiv\sum_k\,v_k^2\,\langle
\phi_k(t)|\hat{\mathscr{Q}}|\phi_k(t)\rangle$ is recorded as a function of time.
A Fourier transform is then performed on $\Delta q(t)\equiv q(t)-q(t=0)$ to 
obtain the strength function corresponding to $\hat{\mathscr{Q}}$:
\begin{equation}
	\label{strengths}
	S(E;\hat{\mathscr{Q}})=
	-\frac{1}{\pi \hbar \epsilon}{\rm Im} \int \Delta q(t)\,dt\,e^{(iE-\Gamma/2) t/\hbar},
\end{equation}
where $\Gamma$ (in MeV) is the smoothing width.

\subsubsection{The energy-weighted sum rules (EWSRs)}
\label{ewsr}

From the strengths, the EWSR can be calculated using 
\begin{equation}
	\label{sum_rule_tdhf}
	m_1 = \int E \times S(E) d E.
\end{equation}
Recently, the EWSR for the density functional theory has been systematically
derived in Refs.~\cite{hino15,hino19}. The EWSRs for the IVD operator has been
shown in Ref.~\cite{shi20}. For the isoscalar operators considered here, the sum
rule using Eq.~(62) of Ref.~\cite{hino19} can be adapted as follows
\begin{equation}
	\label{sum_rule}
	m_1\qty(\hat{\mathscr{Q}}_{\lambda\mu}) = \frac{\hbar^2}{2m}\alpha^2\int d\bm{r}\,
	\qty(\grad{\mathscr{Q}_{\lambda\mu}})^2 \, \rho(\bm{r}),
\end{equation}
where $\rho=\rho_n+\rho_p$, and $m=\frac{1}{2}(m_n+m_p)$.


The EWSR value obtained from Eq.~(\ref{sum_rule}) is related to various
densities of the g.s. Thus, they can be determined rather precisely.  To what
extent the $m_1$ values obtained from TDDFT [Eq.~(\ref{sum_rule_tdhf})] and
Eq.~(\ref{sum_rule}) agree forms a stringent testing ground for the \HIT~code.

\subsection{Numerical solution of the static and time-dependent problems}
\label{fd}

As shown in Eq.~(\ref{hamiltonian}), the s.p. Hamiltonian of the SHF problem
contains systematic derivative operations on the wave functions up to the second
order. This can be seen more specifically in \ref{Dterm}, where the most
complicated operation of $\hat{h}_q$ on a spinor is specifically expanded. In
this section, we present in some detail about the numerical methods used to
solve the DFT (\ref{dft}) and TDDFT (\ref{td}) problems. In
Sec.~\ref{grid_arrangement}, we show the way \HIT~discretizes the 3D Cartesian
space. Section~\ref{fd} presents the way \HIT~evaluates the first and second
derivatives of a function, namely, the finite-difference method used by the
code. Finally, the imaginary time step method \HIT~code adopts to solve the HF
problem is presented in Sec.~\ref{itsm}.

\subsubsection{The grid points arrangement}
\label{grid_arrangement}

The grid points in the present implementation are moved away from the origin of
the simulating box and differs from those of Ref.~\cite{shi18}. Specifically,
in the example of one dimension, instead of using a set of coordinates at 
\begin{equation}
	\label{oldgrid}
[-n_x^{\rm max}, ...,0,1,...+n_x^{\rm max}]\times dx,
\end{equation}
the \HIT~code represents the problem on grid points at the coordinates 
\begin{equation}
	\label{newgrid}
[-n_x^{\rm max}+0.5, ...,-0.5,0.5,...+n_x^{\rm max}-0.5]\times dx,
\end{equation}
where $n_x^{\rm max}$ is an integer enumerating the points at the edge of the
simulating box. $dx$ denotes the grid spacing. Note that the latter choice
(\ref{newgrid}) has an even number of grid points, whereas the former one
(\ref{oldgrid}) has an odd number of grid points. This choice is guided by the
fact that the inclusion of the grid point at the origin of the box results in
numerical problems~\cite{maruhn14}. Using the grid shown in Eq.~(\ref{newgrid}),
the integration can be carried out by summation {\it on} the grid, without the
interpolation as presented in Ref.~\cite{shi18}.

\subsubsection{Derivatives on the grid points}
\label{fd}

\begin{table*}[htb]
	\centering
	\caption{The finite-difference coefficients for the first-order ($c_m$)
	and second-order ($d_m$) derivatives. For the first derivatives, we use
	a seven-point central formula (Eq.~(\ref{first_order})). Whereas for
	the second derivatives, we use a nine-point formula (\ref{second_order}).}
\label{tab_fd}
\begin{tabular}{cccccccccc}
\hline
\hline
	$k$ & $i-4$ & $i-3$ & $i-2$ & $i-1$ & $i$ & $i+1$ & $i+2$ & $i+3$ & $i+4$ \\
       \hline
	$c_m$ & $-$ & $-\frac{1}{60}$ & $+\frac{9}{60}$ & $-\frac{45}{60}$ & 0 &  $+\frac{45}{60}$ &  $-\frac{9}{60}$     
	&   $+\frac{1}{60}$    & $-$  \\
       \hline
	$d_m$ & $-\frac{9}{5040}$ & $+\frac{128}{5040}$ & $-\frac{1008}{5040}$ & $+\frac{8064}{5040}$ &  $-\frac{14350}{5040}$    & 
	$+\frac{8064}{5040}$	&  $-\frac{1008}{5040}$     &   $+\frac{128}{5040}$   & $-\frac{9}{5040}$  \\
\hline
\hline
\end{tabular}
\end{table*}

The \HIT~code uses the finite difference (FD) formulae to approximate the derivatives
of various densities and wave functions. We use the seven- and nine-point formulae 
to approximate the first and second derivatives at $x_i=(i-0.5)\times dx$, respectively:
\begin{subequations}
\begin{equation}
	\label{first_order}
	f'(x_i)\approx \frac{1}{d x}\sum_{k=i-3}^{i+3} c_k f(x_k),
\end{equation}
and
\begin{equation}
	\label{second_order}
	f''(x_i)\approx \frac{1}{dx^2}\sum_{k=i-4}^{i+4} d_k f(x_k).
\end{equation}
\end{subequations}
The $c_k$ and $d_k$ values are listed in table~\ref{tab_fd}.

The values of the function outside the simulating box are unknown. To calculate
the derivatives at the border region, we set the values outside the box to be
zero. That is, we use the Dirichlet boundary condition for solving the static
and time-dependent problems.

Although the FD method is simple, it turns out to be fairly accurate. In
Section~\ref{nonconstraint_result}, we will demonstrate that, from the lighter
($^{48}$Ca) to the heaviest ($^{208}$Pb) doubly magic nuclei, the method
overbinds by only $\sim0.1\%$ of the total binding energy with $d x=1.0$\,fm.
It has to be mentioned that such a simple method provides a base for further
refinements, such as the Lagrangian interpolation method~\cite{ryss15a} or a
Fourier transformation method~\cite{maruhn14,Schuetrumpf2018}.

\subsubsection{Imaginary time step method}
\label{itsm}

To solve the nonlinear HF problem presented in Eq.~(\ref{hamiltonian}), one
can expand the Hamiltonian on a basis and diagonalize it. For the nuclear HF
problem where only a few tens of s.p. levels are occupied, one finds the
imaginary time step method introduced in Ref.~\cite{davies80} to be quite
efficient.

The imaginary time step method includes two steps to advance the variation
process until the convergence of the nonlinear HF problem. First, one
evaluates the following transformed s.p. wave functions
\begin{equation}
	\label{imagiary_time}
	\Phi_k^{(i+1)}(\bm{r}\sigma)=\Bigg(1-\frac{\delta t}{\hbar}\hat{h}\Bigg)\,\phi_k^{(i)}(\bm{r}\sigma).
\end{equation}
The indexes $(i)$ label the iteration number. Note, that $\Phi_k^{(i+1)}$'s
are not orthonormal now. Second, one performs a Gram-Schmidt procedure to
orthonormalize $\Phi_k^{(i+1)}$'s, obtaining the $\phi_k^{(i+1)}$ for the
$(i+1)$-th iteration.

\begin{figure}[tbp]
\centering
\includegraphics[scale=0.10]{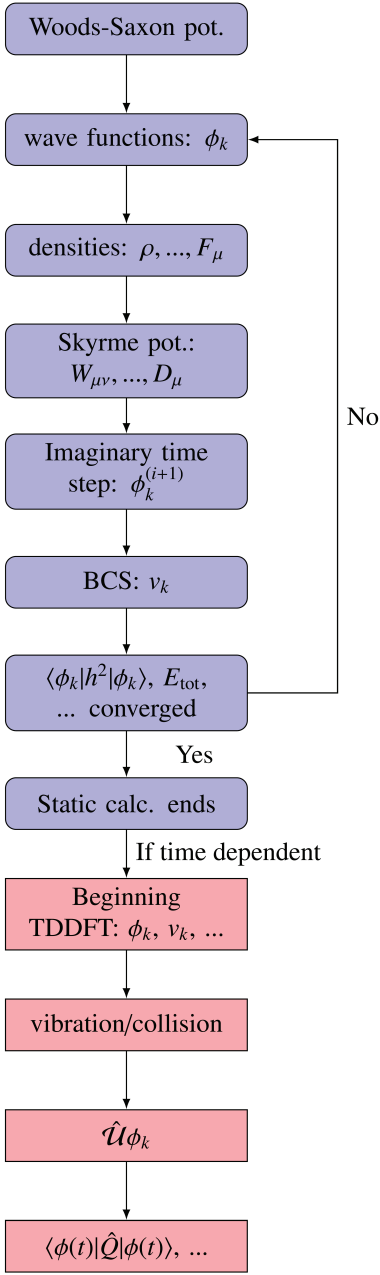}
	\caption{A flowchart demonstrating the main procedures the \HIT~code takes
	to solve the nuclear (TD)DFT problem.}
\label{flowchart}
\end{figure}

We end this section by summarizing the procedures that the \HIT~program adopts
to solve the static and time-dependent problems. In Fig.~\ref{flowchart}, we
provide a flowchart to facilitate the explanation.
\begin{itemize}

	\item[1.] The calculation starts by solving the 3D Schr\"odinger equation
	with a triaxially deformed Woods-Saxon (WS) potential~\footnote{For clarity,
	the WS solution is obtained first with the time-reversal symmetry on a 
	deformed 3D HO basis. The other half of the time-reversed s.p. levels are
	obtained through a time-reversal operation on them and are added to the guessed
	solutions to initiate the HF iteration.}.
	Optionally, the code can read the s.p. wave functions recorded on a file.
	The s.p. wave functions can be solutions to the same static problem but
	with a nearby deformation.

	\item[2.] Retaining the requested number of neutron and proton
		wave functions, \HIT~computes the densities
		(\ref{local_densities}) and potentials (\ref{potentials_full}).

	\item[3.] With the s.p. Hamiltonian, one obtains the s.p. energies and the 
		sum of the fluctuations of each s.p. orbital, weighted by their 
		occupation probabilities~\cite{ryss15a}
		\begin{equation}
			\label{fluct}
			\frac{1}{A}\sum_k v_k^2  \Delta\epsilon_k^2=
			\frac{1}{A}\sum_k v_k^2 \Big(\langle \phi_k | \hat{h}^2 |\phi_k\rangle- \langle \phi_k | \hat{h} |\phi_k\rangle^2 \Big).
                \end{equation}
	\item[4.] The HF iteration starts by evolving the s.p. wave functions
		using an imaginary time step method. For this and the last
		step, one needs to evaluate $\hat{h}\phi_i$. In~\ref{Dterm}, we
		include how this is performed for the most complicated term
		($\bm{D}$-term) in Eq.~(\ref{hamiltonian}).

	\item[5.] If the differences in total energies, weighted fluctuation
		sums (\ref{fluct}), and quadrupole moments between consecutive
		iterations are small enough, the static calculation is
		considered to be converged. \footnote{Note, that these
		calculations often appear converged with the the energy
		differences and the deformations changing quite slowly. In such
		situations, it is important to check if the fluctuation
		(\ref{fluct}) is reduced to a small number
		($<10^{-3}$\,MeV$^2$) for a better-converged solution.} 

	\item[6.] After obtaining a set of orthonormal wave functions upon
		convergence, one performs spatial boost on the s.p.
		wave functions (\ref{boost_isoscalar}-\ref{boost_trans}).

	\item[7.] The time-dependent evolution (\ref{evolve}) is then performed
		on the s.p. wave functions.  The quadrupole moments are recorded
		for the Fourier transformation (\ref{strengths}).

\end{itemize}

\section{Comparisons with other codes}
\label{comparisons}

\subsection{Static cases}

In this section, we present a set of static calculations using the \HIT~code.
For the non-constrained DFT calculations, comparisons are made with the
Sky3D codes (Sec.~\ref{nonconstraint_result}). For the constrained DFT + BCS
calculations in Sec.~\ref{constraint_result}, we use the ALM method to obtain
the results with two deformation points for the $^{110}$Mo nucleus. The
\HIT~results are compared with those of the Ev8 code. For these static
calculations, the T44 EDF is adopted without center-of-mass correction.

\subsubsection{DFT Results without constraint}
\label{nonconstraint_result}

\begin{table*}[htb]
	\centering
	\caption{The total ($E_{\rm tot}$), kinetic ($E_{\rm Kin.}$), Coulomb
	($E_{\rm Coul}$), tensor ($E_{J^2}$), and the Skyrme energies without
	the tensor contributions ($E_{\rm Skyrme}$) of $^{48}$Ca, $^{132}$Sn,
	and $^{132}$Pb calculated with the T44 parameter~\cite{lesinski07}. The
	results calculated using the 
	Sky3D~\cite{maruhn14,Schuetrumpf2018} code are also included. The energies are in MeV.
	The center-of-mass correction~[the (1-1/$A$) factor in Eq.~(\ref{com})]
	is not included in these calculations.}
\label{tab_ca48}
\begin{tabular}{llrrrrr}
\hline\hline
	\multirow{2}{*}{Nuclei}& & \multicolumn{4}{c}{\HIT}  & Sky3D \\
         \cline{3-6}
	& & $dx=1.2$ & $dx=1.0$ & $dx=0.8$ & $dx=0.6$  & $dx=0.6$ \\
       \hline
\multirow{5}{*}{$^{48}$Ca} &$E_{\rm tot}$ & $-$401.034 & $-$400.753  & $-$400.759 & $-$400.816  & $-$400.845  \\
	&$E_{\rm Kin.}$             & 837.996    & 837.251     & 837.028    & 837.075           & 837.103     \\
	&$E_{\rm Coul}$             & 71.156     & 71.066      & 71.035     & 71.029            & 71.024      \\
	&$E_{\rm Skyrme}$           & $-$1314.619& $-$1313.568 & $-$1313.352& $-$1313.461       & $-$1313.518 \\
	&$E_{\rm J^2}$              & 4.434      &    4.498    & 4.529      & 4.542             & 4.545       \\
	\hline
	\multirow{5}{*}{$^{132}$Sn} & $E_{\rm tot}$     & $-$1085.306 & $-$1083.747 & $-$1083.556 & $-$1083.632 & $-$1083.709 \\
				    & $E_{\rm Kin.}$    & 2463.306    & 2460.134    & 2459.179    & 2459.115    & 2458.973   \\
				    & $E_{\rm Coul}$    & 341.977     & 341.706     & 341.602     & 341.578     & 341.556 \\
				    & $E_{\rm Skyrme}$  & $-$3909.377 & $-$3904.614 & $-$3903.469 & $-$3903.498 & $-$3903.420 \\
				    & $E_{\rm J^2}$     & 18.788      & 19.029      & 19.134      & 19.171      & 19.182 \\
	\hline
	\multirow{5}{*}{$^{208}$Pb} & $E_{\rm tot}$ & $-$1619.901 & $-$1617.209  & $-$1616.681 & $-$1616.692   & $-$1616.769 \\
	                            &$E_{\rm Kin.}$ & 3893.725      & 3889.133       & 3887.414      & 3887.170   & 3887.226  \\
	                            &$E_{\rm Coul}$ & 798.737       & 798.150        & 797.913       & 797.862    & 797.809   \\
	                            &$E_{\rm Skyrme}$& $-$6332.832   & $-$6325.205   & $-$6322.822   & $-$6322.571& $-$6322.660 \\
	                            &$E_{\rm J^2}$  & 20.469        &    20.714      & 20.814        & 20.848     & 20.856   \\
\hline\hline
\end{tabular}
\end{table*}

In this section, the ground-state energies of a few spherical nuclei are
calculated using the \HIT~and Sky3D codes. Special attention has been paid to
the convergence of the energies with respect to the grid spacing as well as the
contributions due to the time-even part of the tensor contribution.

Table~\ref{tab_ca48} lists the calculated total energies and the decompositions
of $^{48}$Ca, $^{132}$Sn, and $^{208}$Pb. The tensor contributions are nonzero
with the T44 EDF used for these nuclei. For the \HIT~and Sky3D codes, the sizes
of the simulating boxes have been chosen to be big enough so that the densities
at the edge are lower than $10^{-7}$\,fm$^{-3}$.

First, we discuss the performance with the \HIT~code (the first four columns of
Table~\ref{tab_ca48}). For $^{48}$Ca, the total energy with $dx=1.2$\,fm is
overbound by a few hundreds of keV compared to the results with $dx=0.6$\,fm.
With $dx=1.0$\,fm, the total energies are close to those with $dx=0.6$\,fm
($\sim100$\,keV) for $^{48}$Ca and $^{132}$Sn. For $^{208}$Pb, using
$dx=1.0$\,fm overbinds the converged total energy by $\sim500$\,keV.
Decreasing $dx$ to 0.8 or 0.6\,fm, we see the total energies of these three
nuclei converge rather well, with the total energies changing by $<100$\,keV,
indicating good convergence at $dx=0.8$\,fm for the heavies nucleus $^{208}$Pb.
To summarize, for the lighter or medium heavy nuclei with $A<132$, it is
adequate to use a grid spacing of $dx=1.0$\,fm. Whereas for even heavier nuclei
with $A>208$, it is necessary to use a finer grid spacing of $dx=0.8$\,fm if
one needs better-converged g.s. energies.

We compute the three nuclei using the Sky3D~code, listed in the last column in
Table~\ref{tab_ca48}. For $^{48}$Ca, these two methods are converged/saturated
with respect to the respective model parameters. We notice good agreement among
the two sets of calculated results. Indeed, the differences in total energies
are smaller than 50 keV, which is only $\sim0.02\%$ of the total energy. This
marks the limit of the comparability of these two codes. For heavier nuclei, we
notice that the differences among different codes are about $0.02\%$ of the
respective energies.

It is interesting to note that the {\it kinetic energy} agrees rather well
between \HIT~and Sky3D codes. At $dx=0.6$\,fm, the differences are only a few
tens of keV for the three nuclei. This is impressive because the kinetic energy
depends on the Laplacian operation of the s.p. wave functions. The agreement of
the kinetic energy indicates that the wave functions are rather close between
the two codes.

\subsubsection{Constrained DFT Results}
\label{constraint_result}

\begin{table*}[htb]
	\centering
	\caption{The calculated total energies ($E_{\rm tot}$) and its
	decompositions of the kinetic ($E_{\rm Kin}$), Coulomb ($E_{\rm
	Coul.}$), tensor ($E_{\rm J^2}$) energies, and the Skyrme energies
	minus tensor contributions ($E_{\rm Skyrme}$) of $^{110}$Mo calculated
	using T44 EDF. The pairing energies and
	pairing gaps ($\Delta_{n,p}$) are also included. The first two columns
	correspond to the constraint calculations with the targeting quadrupole
	moments of $(q_1,q_2)=(200,200)$\,fm$^2$. The last two columns
	correspond results with desired $(q_1,q_2)=(600,400)$\,fm$^2$. The
	first and third columns are \HIT~results; the second and fourth columns
	are Ev8$^{\tau}$ results. Both codes use a grid spacing of
	$dx=0.8$\,fm. For the pairing strengths, both codes use
	$V_{n,p}=-500$\,MeV\,fm$^3$. All values are in MeV, except for the
	$q_{1,2}$ values, the latter of which are in fm$^2$.}
\label{tab_mo110}
\begin{tabular}{lrrrr}
\hline\hline
	 & \HIT & Ev8$^{\tau}$ & \HIT & Ev8$^{\tau}$ \\
  \hline
$E_{\rm tot}$            & $-$896.986    & $-$897.111      &  $-$898.041   &  $-$898.136 \\
$E_{\rm Kin.}$           & 1985.011      & 1990.244        &  2000.360     &  2005.029  \\
$E_{\rm Coul.}$          & 252.435       & 252.669         &  250.93       &  251.112   \\
$E_{\rm Skyrme}$         & $-$3122.763   & $-$3127.839     &  $-$3151.848  &  $-$3155.745 \\
$E_{\rm J^2}$            & 5.158         &  5.295         &   9.230        &   9.353   \\
$E_{\rm pair}^n$         & $-$12.613     & $-$13.163      &   $-$4.649       &  $-$5.456 \\
$E_{\rm pair}^p$         & $-$4.214      & $-$4.318       &   $-$2.064       &  $-$2.429 \\
$\Delta_n$               & 1.734         & 1.757          &   1.029        &  1.1075   \\
$\Delta_p$               & 1.109         & 1.119          &   0.774        &  0.8364   \\
$q_1$ (fm$^2$)           & 200.054       & 200.109        &   600.070      &  600.135  \\
$q_2$ (fm$^2$)           & 199.942       & 200.187        &   399.868      &  399.590  \\
\hline\hline
\end{tabular}
\end{table*}

Table~\ref{tab_mo110} lists the results of $^{110}$Mo calculated using two
constraint methods: the \HIT~and Ev8$^{\tau}$ codes~\footnote{The columns with
``Ev8$^{\tau}$'' indicate that the calculations are performed using the Ev8
code~\cite{ryss15a}, except that the kinetic density ($\tau_q$) is calculated
using Eq.~(33) of Ref.~\cite{shi18}. Further, the various energies are
calculated $without$ interpolating on the wave functions or densities. Compared
to the original Ev8 code, this procedure gives kinetic energies ($E_{\rm
Kin.}$) closer to those of the \HIT~and Sky3D codes. The treatment of kinetic
density, similar to Ev8$^{\tau}$, can be found in Ref.~\cite{bonche85} but not
adopted in Ev8.}. Comparing the \HIT~results with the Ev8$^{\tau}$ results, it
can be seen that the total energies agree on the level of 0.1 MeV, although the
kinetic energies differ by about 5 MeV. The differences in the kinetic energies
using different codes are well known~\cite{ryss15a}. Note that the two codes
give similar pairing properties (pairing gaps and energies).

For the constraining calculations in Table~\ref{tab_mo110}, it can be seen that
both codes obtain the desired solution at the desired quadrupole moments. The
calculated $q_{1,2}$ values deviate from the targeted values by only $<0.2$
fm$^2$.

The constrained calculations frequently suffer from divergence. To increase
the chance of reaching convergence, we add a few tips when running the code:
\begin{itemize}

	\item Start the calculation with a set of s.p. wave functions, resulted
		from the WS potential with the $basis$ deformation that
		corresponds to the desired quadrupole moments.

	\item The constraining coefficients (\ref{quad}) of relevant
		constraining terms need to be chosen in such a way that the
		constraint energies (\ref{quad}) are not too high, with $E_{\rm
		constr.}=\sum_{\lambda\mu}C_{\lambda\mu}\qty(q_{\lambda\mu}-q^0_{\lambda\mu})^2<10$\,MeV.

	\item If a constraint calculation is successful, use this set of
		wave functions as a starting point for the constraint
		calculation for a nearby deformation.

        \item Use a large mixture of densities from the last iteration. 
		This will tend to slow down the convergence but increase 
		the chance of convergence.

\end{itemize}

\subsection{Dynamic cases}

\subsubsection{Harmonic Vibrations}

\begin{figure}[tbp]
\centering
\includegraphics[scale=0.5]{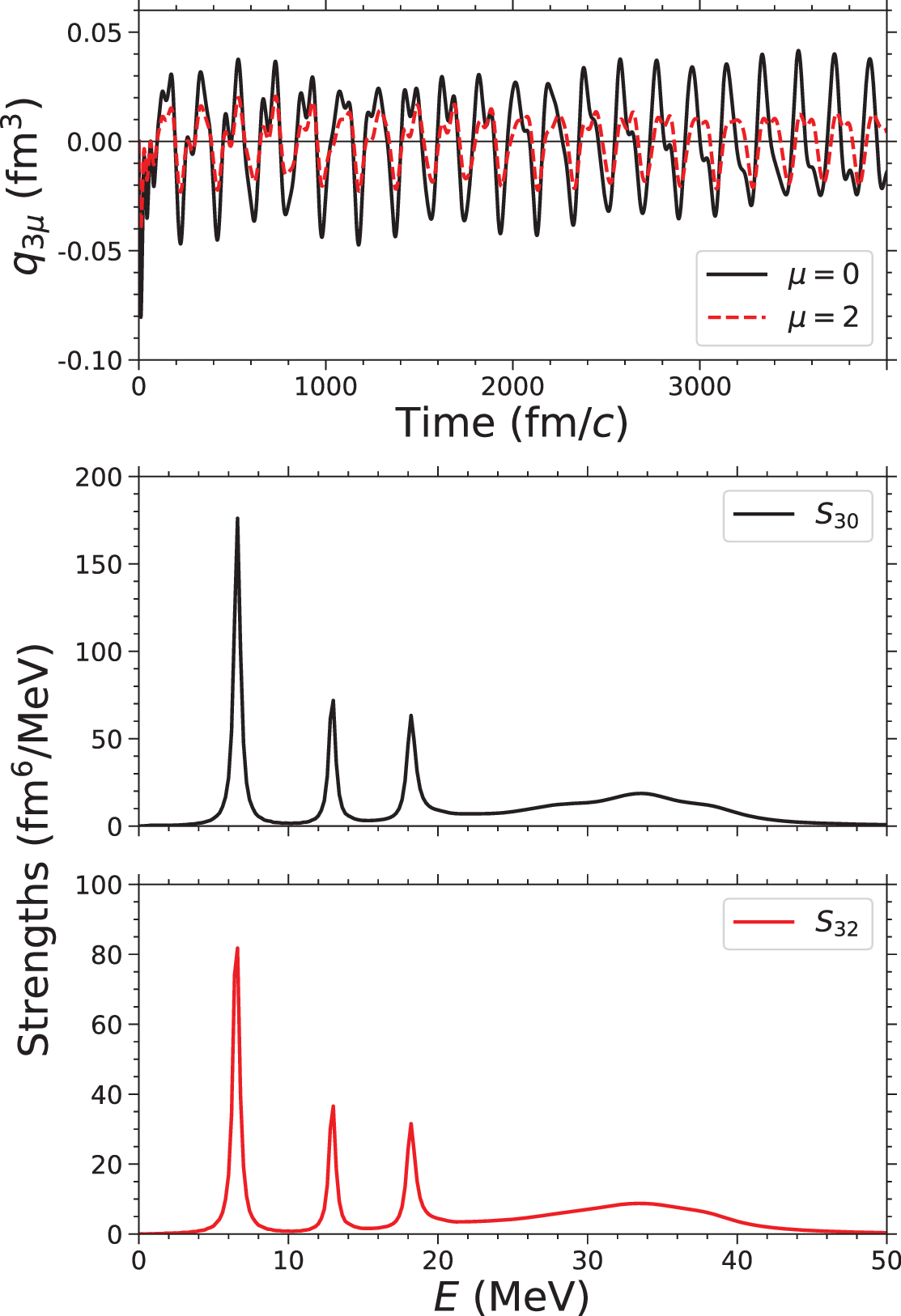}
	\caption{The isoscalar octupole ($\lambda=3$, $\mu=\{0,2\}$) vibrations
	calculations for $^{16}$O. The upper panel shows the octupole moments as
	a function of time. The lower two panels display the strengths functions
	for the $\mu=\{0,2\}$ vibration modes.}
\label{isoscalar_oct_o16}
\end{figure}

Figure~\ref{isoscalar_oct_o16} plots the results for the harmonic isoscalar
octupole vibration of $^{16}$O using the SkM* force.
Figure~\ref{isoscalar_oct_o16}(a) plots the octupole moments up to
$t=4000$\,fm/$c$ after the isoscalar octupole boosts (strength
$\epsilon=0.0001$\,fm$^{-3}$) with $\mu=0, 2$ at $t=0$. It can be seen that the
magnitudes are maintained.

Fourier transformations are performed on the response
functions~(\ref{strengths}) of the octupole moments, obtaining the strengths of
the $q_{30}$ and $q_{32}$ modes, which are displayed in
Figs.~\ref{isoscalar_oct_o16}(b) and (c), respectively. Strong narrow peaks at
6.6, 13.0, and 18.2\,MeV can be seen, suggesting good harmonic vibrations with
phonon energies of $\sim$6.5 MeV. These correspond to the isoscalar $Q_{30}$
and $Q_{32}$ modes. The lowest peaks are the strongest. This seems to be
confirmed by the experimental data, where a few lowest negative-parity states
(with $I^{\pi}=1^-$ and 3$^-$) appear near $E_x\sim6$ to 9\,MeV~\cite{nndc}. In
the energy interval of $E\sim[25,45]$\,MeV, there are broad continuum strengths
corresponding to these octupole modes.

\begin{table*}[htb]
	\centering
	\caption{EWSR of a few doubly magic nuclei obtained for g.s. and the
	TDDFT calculation using SkM* force.}
\label{tab_ewsr}
\begin{tabular}{cccccccc}
\hline
\hline
\multirow{2}{*}{Nuclei}	&	Modes &    ISM            & IVD            & ISQ ($\mu=0$) & ISQ ($\mu=2$) & ISO ($\mu=0$) & ISO($\mu=2$)  \\
	                &	unit  &  fm$^4$MeV  &$e^2$fm$^2$MeV & fm$^4$MeV& fm$^4$MeV& fm$^6$MeV& fm$^6$MeV    \\
\hline
	\multirow{2}{*}{$^{16}$O} & g.s.  &  9908.0   &   75.8  &   492.8   &   246.4  &   11891.8  &  5948.1      \\
				  & TDDFT  &  9863.0  &  75.8   &   491.1   &   245.6  &   11840.9 &  5920.1          \\
\hline
	\multirow{2}{*}{$^{24}$Mg} & g.s.  &  18666.6  &   117.1  &  1161.9  &  347.5  &  34738.4  &  13228.7      \\
				   & TDDFT &  18601.0  & 116.2    &   1158.1 &  346.4  &  34594.2  &  13173.3       \\
\hline
	\multirow{2}{*}{$^{40}$Ca} & g.s.  &  38904.2  &  197.8 &  1934.9  &  967.5  &  67277.6  &  33658.5      \\
				   & TDDFT &  38513.8  &   196.3&  1928.3  &  964.0  &  66997.9  &  33515.0      \\
\hline
	\multirow{2}{*}{$^{48}$Ca} & g.s.  &  50422.9  &  234.9 &  1741.9 &  870.6  &  63975.2  &  32010.6      \\
				   & TDDFT  & 50278.7  &  232.6 &  1737.8 &  867.5  &  63698.6  &  31867.5       \\
\hline
	\multirow{2}{*}{$^{208}$Pb} & g.s.  &  534222.7   &   1031.6 &   16517.8 &  8258.9  &   1373619.5&  687306.1    \\
				    & TDDFT  &  534972.0  &  1023.2  &  16460.8  &  8233.4  &  1365840.0 & 684045.0   \\
\hline
\hline
\end{tabular}
\end{table*}

Table~\ref{tab_ewsr} lists the EWSR calculated using the g.s.
densities~(\ref{sum_rule}) compared with those resulting from a
TDDFT-calculated strengths function~(\ref{sum_rule_tdhf}) for a few nuclei. A
good agreement can be seen for all the modes that allowed for the \HIT~code.

\subsubsection{Heavy-ion Collisions}
\label{nuclear_reaction}

When two $^{16}$O collide, with certain center-of-mass kinetic energy, which is
larger than the Coulomb barrier between them, the two nuclei fuse. With higher
center-of-mass energy, the two nuclei will separate into two. The threshold
energy (the highest center-of-mass kinetic energy at which the fuse happens) is
sensitive to the included Skyrme terms. Consequently, it has been customarily
used as a benchmark when new terms are added to the TDDFT applications for
nuclear reactions~\cite{umar06,stev16}.

Table~\ref{tab_headon} lists the threshold energy for the two head-on (with
$b=0$) colliding $^{16}$O nuclei. Comparing the results calculated using
\HIT~with the results using Sky3D code~\cite{stev16}, we see a systematic
underestimation of the threshold of \HIT~results by $\sim1$ to
2\,MeV~\cite{stev16}. It has to be noted that the differences become smaller
(by $\sim1$\,MeV) if the Sky3D uses similar grid spacings and colliding angles
as those adopted by the \HIT~code.

\begin{table}[htb]
\centering
	\caption{The threshold energies for the head-on
	$^{16}$O-$^{16}$O collision ($b=0$) calculated using the TDDFT
	code \HIT~with different forces. The SLy5T is the force fitted for the
	shell evolutions~\cite{colo07}.}
\label{tab_headon}
\begin{tabular}{lcc}
\hline
\hline
	Force  & Threshold (MeV) \\
\hline
        SkM* (full) & 72 \\
	SLy5T\cite{colo07} & 69 \\
        T12 & 60 \\
	T22 & 64 \\
        T26 & 80 \\
	T44 & 77 \\
	T46 & 85 \\
	T64 & 85 \\
	T66 & 91 \\
\hline
\hline
\end{tabular}
\end{table}

\section{Input description}
\label{inputs}

The input parameters are included in two files. The header file
\verb+headers.tensor.108.h+ contains those parameters fixed at the compilation
stage. While the input file \verb+input_118.dat+ contains parameters sent to
the executable file.

\subsection{Header file}

The parameters inside the header file \verb+headers.tensor.108.h+ 
are fixed at the compilation stage. They can stay unchanged for one
mode of calculation.

\begin{itemize}
\item \verb+HOshell_+: the largest $n$ for the Hermite functions used to 
	expand the WS Hamiltonian.
\item \verb+Spcing+: the grid spacing ($dx,dy,dz$) fm for both the HF and TDDFT calculations.
\item \verb+delta_t+: the step size of the imaginary time [$\delta t$ in
	Eq.~(\ref{imagiary_time})] in units of 10$^{-22}$\,s. A large $\delta
	t$ leads to divergence of the HF problem since we expand the
	exponent up to the first order in Eq.~(\ref{imagiary_time}).
	With $dx=1.0$\,fm, it is found that $\delta t=0.04$
	leads to divergence. We use a much smaller $\delta t=0.005$
	to ensure convergence for grid spacings ranging from 0.6 to 1.2 fm.
	Static calculations with smaller grid spacings require even smaller $\delta t$
	to achieve convergence.
\item \verb+Rab+, \verb+dRab+, and \verb+eta_0+: the $R$, $\Delta r$, 
	and $\eta_0$ parameters for the ABC (\ref{eq:ABC}).
\item \verb+Nx_max+: this parameter specifies the largest $N$ quantum number 
	in the $x$ direction of the 3D HO wave functions expanding the WS Hamiltonian.
	The actual largest value is \verb+Nx_max+-1.
\item \verb+Ny_max+, \verb+Nz_max+: similar to \verb+Nx_max+, except for the $y$, and $z$ directions. 
\item \verb+N+: the number of ($n_x$, $n_y$, $n_z$) points for the WS potential.
\item \verb+beta2_ws+: the $\beta_2$ deformation of the WS potential.
\item \verb+gamma_ws+: the $\gamma$ deformation of the WS potential.
\item \verb+max_x+, \verb+xo+: \verb+xo+ and \verb+max_x-1+ specify the left-most
	and right-most points in the $x$-axis. Specifically, the coordinate 
	on the left-most is \verb+(xo-0.5)+$\times$\verb+Spcing+. On the right-most
	is \verb+(max_x-1.5)+$\times$\verb+Spcing+.
\item \verb+max_y+, \verb+yo+: similar to \verb+max_x+ and \verb+xo+, except in the $y$ direction.
\item \verb+max_z+, \verb+zo+: similar to \verb+max_x+ and \verb+xo+, except in the $z$ direction.
\end{itemize}

\subsection{Input file}

The data file \verb+input_118.dat+ contains input parameters for the executable
file to read. The program reads the file line by line. The meaning of the lines
is determined by the order they appear in the data file. Thus, no line can be
omitted, and the order of the lines needs to be kept. For most lines, the first
keywords are there to remind the user what they are changing. The program reads
the data after the keywords. This section discusses all the parameters in the
data (\verb+input_118.dat+) file.

\begin{itemize}

\item \verb+NZ+: after this keyword, two real numbers are given to indicate the
	neutron and proton numbers of the nucleus to be calculated. We take an
		integer of the two real numbers. If the resulting integers are
		both even, the TDDFT + BCS calculation will be performed. If
		either or both of the two integers is/are odd, BCS is switched
		off.

\item \verb+N_wf+: two integers are provided to indicate the number of wave
	functions for the neutrons and protons, respectively. They should be
		larger than the neutron and proton numbers, respectively.

\item \verb+Force+: a string is provided to indicate the force name.

\item \verb+pair_type+: an integer is provided. ``0'' for delta force;
	``1'' for mixed pairing~(\ref{pairing_strengths}).

\item \verb+v0n+: the neutron pairing strength~(\ref{pairing_strengths}). It is in the unit of MeV fm$^3$

\item \verb+v0p+: similar to \verb+v0n+, except for proton pairing strength.

\item \verb+tensor_on+: an integer is provided after this keyword. If it is
	``0'', then the tensor contribution in Eq.~(\ref{EDF1}) is ignored.
		Specifically, we set $b_{14,15,16,17}$ to be zero. 
		If it is ``1'', then
		the tensor contribution in the form of Eq.~(\ref{hamiltonian})
		is included. If it is ``2'', then the last term in
		Eq.~(\ref{hamiltonian}) is calculated using
		Eq.~(\ref{barton18}) from Ref.~\cite{bart17}, see \ref{Dterm}.

\item \verb+st_on+: an integer is provided. If it is ``0'', the terms inside
	the same bracket as $\bm{s}\vdot\bm{T}$ in the EDF (\ref{EDF1}) are not
		included. Specifically, we set $b_{14}$ and $b_{15}$ to be
		zero. If it is ``1'', those terms are included.

\item \verb+sf_on+: similar to \verb+st_on+, except for the terms inside the
	same bracket as $\bm{s}\vdot\bm{F}$ in the EDF~(\ref{EDF1}).

\item the line after \verb+sf_on+ contains an integer followed by a file name.
	If the integer is ``1'', the static run will be restarted from the wave
		functions in the file with the filename provided after the
		``1'' in this line. If the integer is ``0'', the run will start
		without reading information from another file. Instead, it will
		start by calculating the WS wave functions.

\item this line, which is before \verb+vibration+, contains an integer followed
	by a file name. If the integer is ``1'', the s.p. wave functions of the
		current static run will be written in the file with the file
		name provided after the ``1'' in this line. If the integer is
		``0'', the information will not be saved.

\item \verb+vibration+: an integer is provided after this keyword. If it is
	``1'', the time-dependent calculation will be performed for a harmonic
		vibration [isoscalar boost (\ref{boost_isoscalar})]. If it is
		``2'', the isovector boost (\ref{boost_isovector}) will be
		performed. If it is ``0'', the time-dependent vibration
		calculation will not be performed. The mode is specified with
		the \verb+ExcMod+ keyword.

\item \verb+ExcMod+: an integer is provided after this keyword to select the
	vibration mode to excite. The integers ``0'', ``1'', ``20'', ``21'',
		``22'', ``30'', and ``32'' are available to excite the
		isoscalar monopole, isovector dipole, isoscalar/isovector
		quadrupole ($\mu=0, 2$), and octupole ($\mu=0, 2$) vibrations,
		respectively. For the last five modes, to specify the isoscalar
		or isovector vibration, one can refer to the \verb+vibration+
		keyword.

\item \verb+kick_xyz+: an integer is provided after this keyword. If ``1'' is
	provided, the isovector dipole vibration is in the direction of
	$(\hat{\bm{x}},\hat{\bm{y}},\hat{\bm{z}})$. If ``0'' is provided, the
	isovector vibration is only in the $z$ direction. This keyword does not
	concern other modes except for the isovector dipole vibration.

\item \verb+epsilon_monopole+: the boost parameter $\epsilon$ (in fm$^{-2}$)
	for the isoscalar monopole vibration.

\item \verb+epsilon_quadrupole+: the boost parameter $\epsilon$ (in fm$^{-2}$)
	for the isoscalar/isovector quadrupole vibrations. 

\item \verb+epsilon_dipole+: the boost parameter $\epsilon$ (in fm$^{-1}$) for
	the isovector dipole vibration. 

\item \verb+epsilon_octupole+: the boost parameter $\epsilon$ (in fm$^{-3}$)
	for the isoscalar/isovector octupole vibrations. 

\item \verb+collision+: an integer is provided after the keyword. 
	\begin{itemize}
	\item	If ``0'' is provided, the collision calculation will not be performed.
	\item	If ``1'' is provided, the calculation will compute a testing case where the
		nucleus from the static run will be replaced at a new position (see \verb+frag_one+)
		and boosted across the box.
	\item	If ``2'' is provided, the same two nuclei calculated from the static run will
		be duplicated to two positions (see \verb+frag_one+ and \verb+frag_two+ to
		specify positions) and boosted.
	\item	If ``3'' is provided, the first nucleus will be from the current static run.
		The second colliding nucleus will be read from a file specified in the next line.
	\end{itemize}

\item the line after keyword \verb+collision+ contains an integer and a file
	name. If the previous line \verb+collision+ is ``3'', then the integer
	in this line has to be ``1''.  Meanwhile, the input with the given file
	name has to be supplied. If the integer in the previous line is not
	``3'', then the integer of this line can be ``0''. In this case, the
	file name of this line is ignored.

\item \verb+E_cm+: the total kinetic energy plus the Coulomb energy for the two
	colliding nuclei. It is in the unit of MeV.

\item \verb+frag_one+: three real numbers are provided to indicate the position
	($x_1,y_1,z_1$) of the first nucleus. They are in the unit of fm.

\item \verb+frag_two+: the same as \verb+frag_one+, except for the position of
	the second nucleus.

\item \verb+TaylorExp+: an integer number, ``4'' or ``6'', is provided to
	specify the order of the Taylor expansion $m$ in Eq.~(\ref{expansion}).

\item \verb+delta_t+: a real number (in the unit of fm/$c$) is provided to
	specify the time step for TDDFT calculations in Eq.~(\ref{expansion}).
	This is to be distinguished from the time step in the imaginary time
	step method, \verb+delta_t+, which is in the header file.

\item \verb+max_ite+: an integer is provided to specify the maximum iteration
	number for the static calculation.

\item \verb+itmax+: an integer is provided to specify the largest number of
	time steps for the TDDFT calculations.

\item \verb+nprint+: an integer is provided. The report for the static
	calculation is printed every \verb+nprint+ iterations. The table for
	dynamic information is printed every 20$\times$\verb+nprint+ time
	steps.

\item \verb+threshold+: a small real number (close to zero) is provided to
	specify the threshold energy (in MeV). If the energy difference between
	this iteration and the last iteration is smaller than it for five
	consecutive iterations, then the static calculation is terminated if
	other thresholds are also reached, see below.

\item \verb+threshold_q1t+: a threshold value for $q_1$ (in fm$^2$) is provided
	after the keyword. If the difference in $q_1$ between the two
	consecutive iterations is smaller than it, the convergence is reached.

\item \verb+threshold_q2t+: similar to \verb+threshold_q1t+ but for the $q_2$
	value.

\item \verb+threshold_disp+: similar to \verb+threshold_q1t+ but for the
	threshold of fluctuation (\ref{fluct}).

\item \verb+alpha_mix+: a real coefficient between 0 and 1 is provided to
	multiply with the densities from the last iteration. Only
	\verb+(1-alpha_mix)+ of the current iteration is taken to the next
	iteration.

\item \verb+nfree+: an integer is provided. After \verb+nfree+ number of
	iterations, the constraints on the $q_1$ and $q_2$ quadrupole moments
	will be removed.

\item \verb+C1t+: two real numbers need to be provided. The first one is the
	quadratic constaint coefficient (in MeV fm$^{-4}$) in Eq. (\ref{quad})
	for the $q_1$ quadrupole moment defined in Eq.~(\ref{q1_q2}). The
	second real number specifies the desired $q_1$ value (in fm$^2$).

\item \verb+C2t+: the same as \verb+C1t+, except for the $q_2$ values defined
	in Eq.~(\ref{q1_q2}).

\item \verb+C30t+: the same as \verb+C1t+, except for $q_{30}$ (in fm$^3$) defined in
	Eq.~(\ref{q30}). It is in the unit of MeV fm$^{-6}$.

\item \verb+C31t+: the same as \verb+C30t+, except for $q_{31}$.

\item \verb+C32t+: the same as \verb+C30t+, except for $q_{32}$.

\item \verb+C33t+: the same as \verb+C30t+, except for $q_{33}$.

\item \verb+C10x+: three real numbers need to be provided after this keyword.
	The first one is the quadratic constraint coefficient (in MeV
	fm$^{-2}$) for the $q_x$ moment~(\ref{qxyz}). The second and the third
	are the desired $q_x$ value (in fm) one aims to calculate for neutrons
	and protons, respectively.

\item \verb+C10y+: the same as \verb+C10x+, except for the $q_y$ moment.

\item \verb+C10z+: the same as \verb+C10x+, except for the $q_z$ moment.

\item \verb+Cxy+: similar to \verb+C10x+, except for the $q_{xy}$
	moment~(\ref{qxy}). The coefficient is in MeV fm$^{-4}$, and the
	$q_{xy}$ moments are in fm$^2$.

\item \verb+Cyz+: the same as \verb+Cxy+, except for the $q_{yz}$ moment.

\item \verb+Czx+: the same as \verb+Cxy+, except for the $q_{xz}$ moment.

\item \verb+epscst+: a real number ranging from 0 to 1.0 is specified. It is
	to be multiplied by the updated $L$ [$\epsilon_{\rm constr.}$ in
	Eq.~(\ref{L_update})]. \verb+epscst+ is for $q_1$ and $q_2$ updates.

\item \verb+epscst_q3+: similar to \verb+epscst+, except for updates on
	octupole moments.

\item \verb+epscst_cent+: similar to \verb+epscst+, except for updates on the
	$q_{x,y,z}$ moments.

\item \verb+epscst_refl+: similar to \verb+epscst+, except for updates on the
	$q_{xy,yz,zx}$ moments.

\item \verb+ral+: a real number ranging from 0 to 1.0 is specified. It is to
	be multiplied by the potentials (\ref{alm_pot}) due to the linear
	constraints on the moments that are intrinsic deformation, $q_1$,
	$q_2$, $q_{30}$, and $q_{32}$.

\item \verb+rall+: a real number is provided after the keyword, which is
	similar to \verb+ral+ except for the moments that are responsible for
	the center of mass and the orientation of the nucleus.

\item  in the line after \verb+rall+, a file name is provided to record the
	multipole moments as a function of time. At the end of this file, the
	strengths associated with this vibration mode are saved as a function
	of energy.

\item the last line of the data file contains a file name to save the various
	energies originating from the energy functional (\ref{EDF1}) as
	functions of time.

\end{itemize}

\section{Acknowledgments}

The current work is supported by National Natural Science Foundation of China (Grant No. 12075068 and No. 11705038),
the UK STFC under grant No. ST/P005314/1, ST/V001108/1,
JSPS KAKENHI Grant No. 16K17680, and No. 20K03964.

\appendix

\section{Relationships between $\{t,x\}$ and $\{b\}$ coefficients}
\label{coefficients}

In this section, we connect the $\{b\}$ coefficients in Eq.~(\ref{Etot}) with
the $\{t,x\}$ coefficients used to define the Skyrme force~\cite{skyrme58}. In
the \HIT~code, the interconnection between $\{b\}$ and ${t,x}$ coefficients are
mediated by the $\{a\}$ coefficients that appeared in Ref.~\cite{bart17}.

Specifically, the ${t,x}$ coefficients are first transformed to $\{a\}$ coefficents:
\begin{align*}
	a_1&=\frac{t_0}{2}\qty(1+\frac{x_0}{2}),  &                	a_2&=-\frac{t_0}{2}\qty(x_0+\frac{1}{2}),    \\
	a_3&=\frac{t_0x_0}{4},                    &                	a_4&=-\frac{t_0}{4},                     \\
	a_5&=\frac{1}{4}\qty(t_1+t_2+\frac{t_1x_1+t_2x_2}{2}),&    	a_6&=\frac{1}{16}\qty(t_2-3t_1-\frac{3t_1x_1}{2}+\frac{t_2x_2}{2}),  \\
	a_7&=\frac{3}{2}\qty(t_2x_2-t_1x_1),      &                	a_8&=\frac{1}{32}\qty(t_2+3t_1),    \\
	a_9&=\frac{1}{16}\qty(3t_1x_1+t_2x_2), &                   	a_{10}&=\frac{1}{8}\qty(t_1x_1+t_2x_2),    \\
	a_{11}&=\frac{1}{8}\qty(t_2-t_1),       &                  	a_{12}&=\frac{1}{4}\qty(t_2x_2-t_1x_1),   \\
	a_{13}&=\frac{t_3}{12}\qty(1+\frac{x_3}{2}),&              	a_{14}&=-\frac{t_3}{12}\qty(x_3+\frac{1}{2}),   \\
	a_{15}&=\frac{t_3x_3}{12},                  &              	a_{16}&=-\frac{t_3}{12},     \\
	a_{17}&=\frac{W_0}{2},                      &              	a_{18}&=\frac{W_0}{2}.                                                  
\end{align*}
For tensor part of the Skyrme EDF, the relations between $\{a\}$ 
and $\{t,x\}$ coefficients are
\begin{align*}
	a_{19} &= \frac{3}{16}\qty(3t_e-t_o),     &     a_{20} &= -\frac{3}{16}\qty(3t_e+t_o),  \\
	a_{21} &= -\frac{1}{4}\qty(t_e+t_o),     &      a_{22} &= \frac{1}{4}\qty(t_e-t_o),     \\
	a_{23} &= \frac{3}{4}\qty(t_e+t_o),      &      a_{24} &= -\frac{3}{4}\qty(t_e-t_o),    \\
	a_{25} &= \frac{1}{16}\qty(3t_e-t_o),     &     a_{26} &= -\frac{1}{16}\qty(3t_e+t_o).
\end{align*}

The ${b}$ coefficients are related to the ${a}$ coefficients through
\begin{align*}
	b_1 &=a_1,                    &   b_2 &= a_2,                                        \\
	b_3 &=a_5,                    &   b_4 &= a_{11}+a_{12},                              \\
	b_5 &=a_6,                    &   b_6 &= a_8+a_9,                                    \\
	b_7 &=a_{13},                 &   b_8 &= a_{14},                                     \\
	b_9 &=a_{17},                 &   b_{9q} &= a_{18},                                  \\
	b_{10} &= a_3,                &   b_{11} &= a_4,                                     \\
	b_{12} &= a_{15},             &   b_{13} &= a_{16},                                  \\
	b_{14} &= -a_{21}-a_{10},     &   b_{15} &=-a_{22}-a_{11},                           \\
	b_{16} &= -\frac{1}{2}a_{23}, &   b_{17} &=-\frac{1}{2}a_{24},                       \\
	b_{18} &= a_{25},             &   b_{19} &= a_{26},                                  \\
	b_{20} &= a_{19},             &   b_{21} &= a_{20}.                                  \\
\end{align*}

Note that for $b_{14}$ and $b_{15}$, contributions from $t_1$ and $t_2$ can be
seen (through $a_{10}$ and $a_{11}$)~\cite{bart17}. This means that even if the
tensor force is ignored ($t_o=t_e=0$), contributions due to $\bm{s}\vdot\bm{T}$
terms in Eq.~(\ref{EDF1}) appear in the Skyrme EDF. Therefore, ideally,
\verb+st_on+ should be turned on even for the EDFs without tensor interaction.
However, only a few parameter sets include the $\bm{s}\vdot\bm{T}$ terms in the
fitting process. The common practice has been to ignore these complicated
terms. Consequently, the users are advised to switch them off to save running
time, especially for the EDFs without tensor interactions included in the
fitting procedure.

\section{The last term in Eq.~(\ref{hamiltonian})}
\label{Dterm}

Now, we specifically expand the last two terms in Eq.~(\ref{hamiltonian}):
\begin{align}
&-\grad\vdot\qty[\hat{\bm{\sigma}}\vdot\bm{C}]\grad\mqty(\phi_{\uparrow} \\ \phi_{\downarrow}) \nonumber \\
&=-\qty[\grad\qty(\hat{\bm{\sigma}}\vdot\bm{C})]\vdot\grad\mqty(\phi_{\uparrow} \\ \phi_{\downarrow}) -\qty(\hat{\bm{\sigma}}\vdot\bm{C})\nabla^2\mqty(\phi_{\uparrow} \\ \phi_{\downarrow})
\end{align}
and
\begin{align}
	\label{D_term}
&-\grad\vdot\bm{D}\hat{\bm{\sigma}}\vdot\grad\mqty(\phi_{\uparrow} \\ \phi_{\downarrow}) \nonumber \\
	& =-\qty(\grad\vdot\bm{D})\qty[\hat{\sigma}_x\grad_x\mqty(\phi_{\uparrow} \\ \phi_{\downarrow})+\hat{\sigma}_y\grad_y\mqty(\phi_{\uparrow} \\ \phi_{\downarrow})
	+\hat{\sigma}_z\grad_z\mqty(\phi_{\uparrow} \\ \phi_{\downarrow})] \nonumber \\
	&\quad -\bm{D}\vdot\grad\qty[\qty(\hat{\bm{\sigma}}\vdot\grad) \mqty(\phi_{\uparrow} \\ \phi_{\downarrow}) ].
\end{align}

Another form of the $b_{16}, b_{17}$ term is obtained in Barton's thesis~\cite{bart17} [Eq. (2.4.8)], which reads
\begin{alignat}{4}
\label{barton18}
-\frac{1}{2}\sum_{\mu\nu}\hat{\sigma}_{\nu}\qty[(\nabla_{\nu}D_{q,\mu})\nabla_{\mu} + 2D_{q,\mu}\nabla_{\nu}\nabla_{\mu}+(\nabla_{\mu}D_{q,\mu})\nabla_{\nu}].
\end{alignat}

\section{Computing times}

\begin{table*}[htb]
	\centering
	\caption{The computing time (in hours) cost for 1000 static iterations
	and 1000 time steps in dynamic calculations. The numbers of wave functions
	for the protons and neutrons are the same as the respective proton and 
	neutron numbers. The boxes have $dx=0.8$\,fm. For $^{16}$O, $^{48}$Ca,
	and $^{132}$Sn, the cubic boxes are used with sizes [$-$11.6,11.6] fm, 
	[$-$13.2,13.2] fm, and [$-$16.4,16.4] fm, respectively.
	The dynamic calculations adopt a Taylor expansion of order $m=4$.
	The tests are performed on an Intel(R) Xeon(R) Gold 5218R processor with 80 cores.}
\label{tab_computing_time}
\begin{tabular}{lrrrr}
\hline\hline
	& \multicolumn{2}{c}{Static}            & \multicolumn{2}{c}{Dynamic} \\
	& \multicolumn{2}{c}{(1000 iterations)} & \multicolumn{2}{c}{(1000 time steps)} \\
	\cline{2-3} \cline{4-5}
Nuclei  & Sequential & OpenMP & Sequential & OpenMP \\
  \hline
	$^{16}$O            &  0.478  &  0.436  &  1.656  & 1.408  \\
	$^{48}$Ca           &  1.939  &  1.556  &  6.496  & 5.172  \\
	$^{132}$Sn          &  10.766 &  7.654  &  33.322 & 25.632  \\
\hline\hline
\end{tabular}
\end{table*}

Table~\ref{tab_computing_time} summarizes the computing times for the two
typical application cases. Focusing on the sequential tests, the time used for
1000 time advances is about three times those used on 1000 static iterations.
It has to be mentioned that the time used is much longer compared to the Ev8
and Sky3D codes, which are written in Fortran.

For the orthonormalization and Laplacian (\ref{second_order}) operations over
the s.p. wave functions, we add a simple OpenMP parallelization.
Table~\ref{tab_computing_time} compares the time used by the code with OpenMP
switched on with those sequential results. We see a significant shortening of
the time used starting from medium-heavy nucleu $^{132}$Sn. This is due to the
long time the code spends on the orthonormalization and Laplacian operations
for the heavier nuclei. Future extensions of the code will include a GPU
parallelization for these computationally expensive parts.



\bibliographystyle{elsarticle-num}







\end{document}